\documentclass[11pt, twoside, table]{predoc2}

\usepackage[utf8]{inputenc}
\usepackage[T1]{fontenc}
\usepackage{newtxtext} 
\usepackage[scaled=.95]{cabin} 
\usepackage[varqu,varl]{inconsolata} 
\usepackage{amsmath}
\usepackage{enumitem}
\usepackage{tabularx}
\usepackage[varg]{newtxmath}
\usepackage{changepage}
\usepackage{hyperref}
\usepackage[colorlinks=true,
    linkcolor=red,
    filecolor=black,
    citecolor=black,      
    urlcolor=black]{hyperref}       
\usepackage{url}            
\usepackage{graphicx}
\usepackage{subcaption} 
\usepackage{float}
\usepackage{booktabs}       
\usepackage{amsfonts}       
\usepackage{nicefrac}       
\usepackage{microtype}      
\usepackage{xcolor}         
\usepackage{amsmath} 
\usepackage{bm}
\usepackage{multirow}

\usepackage{booktabs}
\newtheorem*{definition}{Definition}
\usepackage{comment}

\newcommand{\x}{\mathrm{x}}
\renewcommand{\bm}[1]{#1} 

\newcommand{\rep}{{\mathrm{rep}}}
\newcommand{\loo}{{\mathrm{loo}}}
\newcommand{\post}{{\mathrm{post}}}
\DeclareMathOperator{\Uniform}{\mathrm{Uniform}}
\DeclareMathOperator{\Beta}{\mathrm{Beta}}
\DeclareMathOperator{\Bin}{\mathrm{Bin}}
\DeclareMathOperator{\IF}{\bf{IF}}

\title{LOO-PIT predictive model checking}
\author[1]{Herman Tesso}
\affil[1]{Aalto University, Finland}
\author[2]{Aki Vehtari}
\affil[2]{ELLIS Institute Finland, Aalto University, Finland}

\makeatletter
\def\runauthor{Tesso and Vehtari}
\def\runtitle{LOO-PIT}
\makeatother

\begin{document}

\maketitle

\begin{abstract}
We consider predictive checking for Bayesian model assessment using leave-one-out probability integral transform (LOO-PIT). LOO-PIT values are conditional cumulative predictive probabilities given LOO predictive distributions and corresponding left out observations.  For a well-calibrated model, LOO-PIT values should be near uniformly distributed, but in the finite sample case they are not independent, due to LOO predictive distributions being determined by nearly the same data (all but one observation). We prove that this dependency is non-negligible in the finite case and depends on model complexity. We propose three testing procedures that can be used for continuous and discrete dependent uniform values. We also propose an automated graphical method for visualizing local
departures from the null. Extensive numerical experiments on simulated and real datasets demonstrate that the proposed tests perform (at least) competitively against existing methods aimed at mitigating the effect of data-induced dependence in the PITs and have much better power than standard independence-based uniformity tests that often lead to lower than expected rejection rate. \\

\end{abstract}
\thispagestyle{empty}
\section{Introduction}

\label{sec:introduction}

Responsible Bayesian data analysis must assess the plausibility of the entertained model. That is, apart from rare situations, the Bayesian model specification---data model and prior---is unlikely to be exactly the same as the true data generating process. Rather the questions are: ``is the posited model adequate?'', if not, ``which aspects of the model need improvement?''. These questions bring into sharp focus the problem of \emph{model criticism} \citep{box1980sampling,edition2013bayesian,blei2014build}. Although many approaches for criticizing models have been developed, the most prevalent in the Bayesian modelling context are those that adopt a predictive perspective to assess the model's inadequacies \citep{guttman1967use, rubin1984bayesianly, gelman1996posterior}. They involve comparing the model’s predictive simulations---draws from the predictive distribution---with the actual observed data. 

We focus on pointwise predictive check techniques that compare the observed data to the collection of univariate conditional distributions. Across both Bayesian and non-Bayesian literatures (e.g., probabilistic forecasts), pointwise checks utilize the \textit{probability integral transformation} (PIT) to recast model evaluation into a problem of testing for uniformity. PIT corresponds to the model’s predictive univariate cumulative distribution function (CDF) evaluated given the observation, resulting in values that, for a well-calibrated model, should be near uniformly distributed between $0$ and $1$. In this realm, the leave-one-out (LOO) cross-validation approach termed LOO-PIT \citep{gelfand1992model, edition2013bayesian} has been used as part of the Bayesian workflow \citep{gelman2020bayesian,Gelman-Vehtari-Mcelreath-etal:2026} for model checking. It iteratively holds out each observation, conditioning on the remaining data,
and compares the corresponding LOO predictive distributions to the heldout
observations, by testing for uniformity of LOO-PIT values. 

However, in using pointwise LOO-PIT values for diagnosing Bayesian models,  a subtle but crucial point is that, although individual PIT values are asymptotically marginally uniform  given the true model, they are not jointly independent. 
Even if each observation is not used to update the posterior when making predictions for that observation, each observation is used to update all other LOO posteriors. As a result, the LOO predictive distributions across folds have dependency, which in turn induces dependence among the LOO-PIT values (details in Section~\ref{sec:loo-pit dependence}). Although \citet{gelfand1992model} already recognized this, this aspect has been mostly disregarded in LOO-PIT checking. For instance, the graphical uniformity test developed by \cite{sailynoja2022graphical}---provides simultaneous confidence bands for the whole empirical cumulative distribution function (ECDF) of the PIT values---has been used in the Bayesian workflow \citep{gelman2020bayesian} also for LOO-PIT. The dependency in LOO-PIT values reduces the variability of ECDF compared to independent uniform variables, and the envelope based on the independence assumption is too wide, reducing the test's ability to reveal model miscalibration. More generally, standard uniformity tests that assume independence \citep[e.g.,][]{marhuenda2005comparison} are no longer strictly valid. Considering the advantages of LOO-PIT checking in Bayesian model assessment, there is a clear need for uniformity tests that can explicitly accommodate dependent LOO-PIT values.

\begin{figure}[tp]
    \centering    \includegraphics[width=\linewidth]{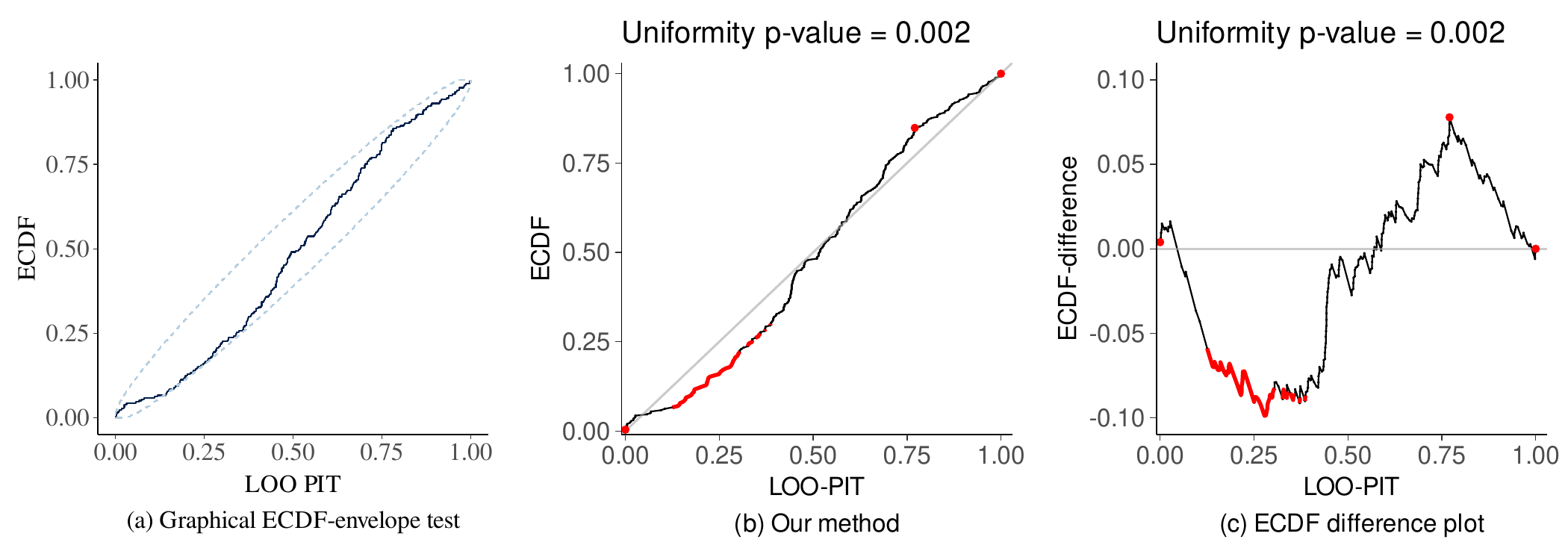}
    \caption{Graphical uniformity checks for LOO-PIT: (a) Method by \citet{sailynoja2022graphical}: Binomial distribution-based 95\% simultaneous confidence intervals for the ECDF values (y-axis). (b) Our method: We aggregate beta distribution-based pointwise tests on quantiles (x-axis) or binomial distribution-based pointwise tests on ranks (scaled y-axis) and provide automated color-coding representations that emphasize problematic parts of the ECDF. (c) Tilted ECDF, $\hat{F}(x)-x$, providing a centered visualization of (b). }
    \label{fig:0}
\end{figure}

We present testing procedures that can handle any dependent uniform values. The methods build on the idea that statistical checks remain valid for each pointwise LOO-PIT value, which can then be combined using a Cauchy aggregation technique for dependent $p$-values \citep[e.g., ][]{liu2020cauchy}. Our contributions include:

\begin{enumerate}[topsep=0pt,partopsep=2pt,itemsep=2pt,parsep=2pt]
    \item We provide theoretical results that clarify how dependence arises in LOO-PIT and posterior PIT values, and how the strength of the correlations relates to model complexity and data size.
    \item We develop uniformity tests that maintain good power performance even under general correlation structures.
    For continuous PIT values we propose \textit{pointwise order tests combination} (POT-C), which aggregates beta-based $p$-values computed from uniform quantiles, and \textit{pointwise inverse-CDF evaluation tests combination} (PIET-C), which combines pointwise $p$-values computed from inverse-CDF evaluations of the uniform sample using a continuous reference distribution. For discrete PIT values we propose \textit{pointwise rank-based individual tests combination} (PRIT-C), which combines binomial-based $p$-values computed from ranks (scaled ECDF).
\item We introduce an efficient graphical procedure based on the proposed tests, which combines color coding with the classical ECDF plot to offer visual insight as to how the observed process deviates from the uniformity assumption. Figure~\ref{fig:0} contrasts our visual approach with that of the ECDF envelope \citep{sailynoja2022graphical}.
\item We perform extensive simulations and real-data experiments to validate the proposed testing procedures and to demonstrate their benefit over closely related diagnostics and tests that assume independence, in terms of test power. The main use case examples in this paper  arise from Bayesian model checking with the LOO-PIT approach, but the proposed methods are more broadly applicable.  
\end{enumerate}

Section~\ref{sec:predictive-checking} reviews predictive checking approaches. Section~\ref{sec:loo-pit dependence} establishes the theoretical basis for how dependence arises in LOO-PIT values and examines its asymptotic behaviour. Section~\ref{sec:dependence-aware} develops the proposed testing procedures, along with the accompanying visualization method. Section~\ref{sec:other-methods} presents alternative methods aimed at addressing data-induced dependence and discusses their connections to LOO-PIT. Section~\ref{sec:experiments} demonstrates PIET-C, POT-C, and PRIT-C empirically on checking complex hierarchical models using synthetic and real data. Section~\ref{sec:discussion} concludes the paper. 

\section{Predictive checking}
\label{sec:predictive-checking}
This section reviews common methods for Bayesian model checking, with a particular emphasis on pointwise predictive checking, including LOO-PIT and posterior PIT. 

\subsection{Posterior predictive checking (PPC)}
Posterior predictive checking compares posterior predictive distributions to observations \citep{rubin1984bayesianly}.

Consider a model of the joint distribution of the data $\bm{y}$ and parameters $\bm{\theta}$,
    $p(\bm{y},\bm{\theta})=p(\bm{y}|\bm{\theta})p(\bm{\theta}),$
where $p(\bm{y}|\bm{\theta})$ is the data model for $y$ and $p(\bm{\theta})$ is the prior distribution for $\theta$. The posterior distribution is $p(\bm{\theta}|\bm{y}^{\mathrm{obs}}) \propto p(\bm{y}^{\mathrm{obs}}|\bm{\theta})p(\bm{\theta})$, where $p(\bm{y}^{\mathrm{obs}}|\bm{\theta})$ is the likelihood given observed data $y^{\mathrm{obs}}$. The posterior predictive distribution is
\begin{equation}
    p(\bm{y}^\rep|\bm{y}^{\mathrm{obs}})=\int p(\bm{y}^\rep|\bm{\theta})p(\bm{\theta}|\bm{y}^{\mathrm{obs}})d\bm{\theta}.
    \label{eq2}
\end{equation}
If the posterior predictive distribution
generates simulated data $\bm{y}^\rep$ that looks like $\bm{y}$, we consider the model to be good. For a selected diagnostic statistic $T(\bm{y})$---function mapping the data space to the real numbers---which is typically chosen to reflect our inferential interests (e.g., $T(y)=\frac{1}{n}\sum_iy_i$), PPC locates $T(\bm{y})$ in the sampling distribution of $T(y^{\text{rep}})$ induced by the posterior predictive distribution. The corresponding tail-area probability is known as the \textit{posterior predictive $p$-value} \citep{meng1994posterior},
 \begin{equation}
      p_{\mathrm{ppc}}=Pr\big\{T(\bm{y}^\rep)\geq T(\bm{y}^{\mathrm{obs}})|\bm{y}^{\mathrm{obs}} \big\}.
       \label{eq3}
\end{equation}
\cite{gelman2004exploratory}, and \cite{gelman1996posterior} discuss also visual PPC approaches, and show how to develop parameter-dependent statistics $T(\bm{y},\bm{\theta})$---termed realised discrepancies. 

Posterior predictive $p$-values do not in general have uniform distributions even given the true model---except in the special case of pivotal discrepancy measures---but instead tend to have distributions more concentrated near $1/2$. This nonuniformity has
been attributed to ``\textit{double use of the data}'' \citep{bayarri2007bayesian, moran2024holdout}, because the data is first used to construct the posterior predictive distribution and then used again in the observed statistic $T(\bm{y})$ to compute the $p$-value.  

To resolve the double use of the data problem, post-hoc adjusted PPCs that are calibrated to have asymptotic uniform null distributions have been proposed \citep{robins2000asymptotic,hjort2006post}. \cite{ moran2024holdout} argued that calibration by itself does not necessarily improve the power of the check and proposed the holdout predictive check (HPC) method, which divides the data in two parts only once, avoiding double use of data and dependency (there is only one fold). 
In the holdout approach the data used to condition the posterior is much smaller than the original data, which may reduce the power to detect model misspecification. \cite{moran2024holdout} demonstrated empirically that HPC achieves higher power than calibrated variants of PPC and remains approximately uniform in finite sample settings. 

We next focus on methods that compare univariate predictive distributions for individual observations $y_i$. That implies checking separate pointwise (conditional on data) predictive distributions against the data $\bm{y}=\{y_i\}_i$ in the sense that under the null model, $y_i$ may be viewed as a random observation from $P(y_i|\bm{y})$.  The baseline for such approaches is the probability integral transformation, which we introduce next.

\subsection{Probability integral transformation}
\label{sec:pit}
The probability integral transform (PIT) states that if \( X \) has CDF \( F_X \), then the random variable \( U = F_X(X) \sim \operatorname{Uniform}(0,1) \). This fundamental result has been facilitating the construction of distribution-free tests for diagnosing statistical models. For instance, PIT 
is used to derive tests of goodness-of-fit of Bayesian models from tests of uniformity \citep{fruiiwirth1996recursive,dawid2012prequential}. 

Let $y_1,....,y_N$ be independent observations from an unknown continuous distribution $G$. We want to know whether these samples could plausibly have been generated from a known distribution $P$ (i.e., if $G=P$) with CDF $F_X$. The PIT of the observed value $y_i$ is simply the value that the CDF attains at the observation,
\begin{equation}
\label{eq11}
    p_i=F_X(y_i)=\int_{-\infty}^{y_i}P(x)dx.
\end{equation}
If  $F_X$ is intractable, numerical integration can still yield a sufficiently accurate approximation of the $p_i$ provided the corresponding PDF $P$ is tractable. If neither the CDF nor the PDF has a closed-form expression, but we can obtain draws from the corresponding distribution, the PIT values can be estimated using the ECDF, $\hat{F}_X$,  provided that an independent comparison sample \( x_1^i, \dots, x_S^i \sim p(x) \) can be drawn for each \( y_i \),
\begin{equation}
    \label{eq12}
    p_i=\hat{F}_X(y_i)=\frac{1}{S}\sum_{s=1}^{S}\mathbb{I}(x_s^{i}\leq y_i).
\end{equation}
Now, given $G=P$, the distribution of the transformed values $p_i$ is consistent with a continuous uniform distribution on the unit interval $(0,1)$ or a discrete uniform distribution on $(0,\frac{1}{S},\cdots,\frac{S-1}{S},1)$ when approximated according to equation Eq.~\eqref{eq12}. In either case, the question of whether the given observations originate from a reference distribution based on PIT reduces to testing for uniformity of the PIT values.
 
\paragraph{Posterior PIT} As a direct application of PIT in the Bayesian framework, consider similar Bayesian model formulation as for PPC, with the pointwise (conditional on the data) predictive distributions, $p(y_i|\bm{y})=\int p(y_i|\bm{\theta})p(\bm{\theta}|\bm{y})d\bm{\theta},$ as references. If $y_i$ are scalars and continuous, using the identity function, \( T(y_i) = y_i\), gives
\begin{equation}
    p_{\mathrm{post},i}=Pr\big\{T(\tilde{y}_i)\leq T(y_i)|\bm{y}\big\}=\int_{-\infty}^{y_i}p(\tilde{y}_i|\bm{y}) d\tilde{y}_i,\quad \forall i.
      \label{eq13_}
\end{equation}
These correspond to the probability integral transformation  of the observations $y_i$ with respect to their corresponding (pointwise) predictive densities. $ p_{\mathrm{post},i}$ are referred to as \textit{posterior PIT} values \citep{edition2013bayesian}. The double use of the data in conditioning the predictive distribution $P(\tilde{y}_i|\bm{y})$ on $y_i$ tends to pull it closer to the observed value $y_i$, causing PIT values to be biased toward $1/2$. Later in Section~\ref{sec:loo-pit dependence} we show theoretically that the posterior PIT values are uniformly distributed only asymptotically.

\subsection{Sequential predictive checking (SPC)} 
PIT has a long history in probabilistic
forecasting across climatology, economics, epidemiology, and ecology \citep[e.g.,][]{ dawid1984present,fruiiwirth1996recursive,gneiting2007probabilistic,Czado-Gneiting-Held:2009}. 
The basic framework assumes a sequentially defined standard model, $M_t$, on sequences of observations, $y=\{y_{_t}\}_{t=1}^{T},$ for which the sampling distribution at each point $y_t$ depends on latent variables $\bm{\theta}_t$. The index $t$ usually refers to time. At time $t$, the history of the series is $H_{t-1}=\{y_1,...,y_{_{t-1}}\}$, the standard model for $y_t$ has the general form $M_t:=p(y_t|\bm{\theta}_t,H_{t-1})\pi(\bm{\theta}_t)$, 
and the corresponding predictive density is
\begin{equation}
    p(y_t|H_{t-1})=\int p(y_t|\bm{\theta}_t,H_{t-1})p(\bm{\theta}_t|H_{t-1})d\bm{\theta}_t.
\end{equation}
The PIT for $y_t$ is then calculated as in Eq.~\eqref{eq13_} using $ P(y_t|H_{t-1})$ as reference. This procedure does not suffer from the double use of data, as the model is tested with an observation not used to update the posterior.
Under the hypothesis of an ideal forecast, these values should at least asymptotically be uniformly distributed.

\subsection{Cross-validation predictive checking}
The sequential approach is natural for time series, but for more general cases we can avoid double use of data by using cross-validation, which iteratively holds out portions of the data, conditioning on the remaining data, and compares the corresponding posterior predictive distribution to the heldout data. Notably, leave-future-out CV \citep{burkner2020approximate} corresponds to sequential predictive checking described earlier.

We focus here on LOO-CV, but cross-validated pointwise predictions can also be made with $K$-fold-CV. 
The benefit of LOO-CV is that each data set used to create the posterior is close to the full data, and thus the bias due to using a smaller data set is minimized. LOO-CV is expected to have smaller variance than $K$-fold-CV when the inference algorithm is stable and utility or cost function is smooth \citep{Arlot-Celisse:2010}, which holds for Bayesian inference and PIT values, too.

\paragraph{LOO-PIT} 
The predictive distribution for $y_i$ is computed with a
model that is fitted to all of the observations except $y_i$ which we denote as $\bm{y}_{-i}$:
$ p(y_i|\bm{y}_{_{-i}})=\int p(y_i|\bm{\theta})p(\bm{\theta}|\bm{y}_{_{-i}})d\bm{\theta}$, 
 and the LOO-PIT value is
\begin{equation}
    p_{\loo,i}=\int_{-\infty}^{y_i}p(\tilde{y}_i|\bm{y}_{_{-i}}) d\tilde{y}_i.
      \label{eq14}
\end{equation} 
Unlike posterior PIT, LOO-PIT avoids double use of the data. The LOO predictive distribution for $y_i$, $p(\tilde{y}_i|\bm{y}_{_{-i}})$, is not pulled towards $y_i$, so the corresponding PIT values tend to spread out more evenly compared to \textit{posterior PIT}. 

When using the Markov chain Monte Carlo (MCMC) methods, sampling from each LOO posterior $p(\bm{\theta}|\bm{y}_{_{-i}})$ would normally take $n$ times the time of sampling from the full posterior $p(\bm{\theta}|\bm{y})$. Instead, importance sampling with the full data posterior as the proposal can be used \citep{gelfand1992model}.  Pareto smoothed importance sampling \citep{vehtari2017practical,Vehtari-Simpson-Gelman-etal:2024} provides more stable computation and self-diagnostic for importance sampling based LOO computation. 

\paragraph{Implication of overlapping (LOO) folds and posterior averaging} In finite sample settings, exact $i.i.d$ uniformity of the LOO-PIT values is not expected even given the true model, since the predictive distribution is obtained by integrating over the posterior. For example, if the true distribution is normal and the model has unknown location and scale, then the predictive distribution is Student's $t$. Asymptotically this $t$-distribution converges towards normal distribution. In addition, since each LOO predictive distribution is determined by nearly the same data (all but one observation), the PIT computation of Eq.~\eqref{eq14} ends up using the whole data $\bm{y}=(\bm{y}_{-i},y_i)$. Therefore, the LOO-PIT values are dependent, as already commented by \cite{gelfand1992model}. The use of uniformity tests that assume independence would lead to lower than expected rejection rate. Asymptotically, the dependence effect vanishes, and the LOO-PIT values should be uniformly distributed. This is examined in more detail in the following section.

\section{LOO-PIT dependence and limiting behaviour theory}
\label{sec:loo-pit dependence}
This section lays out the theoretical basis of PIT dependency, grounded in asymptotic considerations under standard regularity conditions. 

Let $\mathcal{X},\mathcal{Y}$ be non-empty sets. Denote $P_0$ a distribution on $\mathcal{X}\times\mathcal{Y}\subset \mathbb{R}^d\times \mathbb{R}$. Suppose we have data $\mathcal{D}_n
= \left\{ (\mathrm{x}_i, y_i) \;\middle|\; (\mathrm{x}_i, y_i) \in \mathcal{X} \times \mathcal{Y} \right\}_{i=1}^n$ generated i.i.d. from $P_0$. Further, let $p_{\theta}=:p(y|\x,\bm{\theta})$ be a well-specified parametric model with prior measure $\pi(\bm{\theta})$ on the parameter set $\Theta \subset \mathbb{R}^{p}$, and $\bm{\theta}_0$ be the true parameter. Denote the  posterior $\pi(\bm{\theta}|\mathcal{D}_n)$ and the predictive CDF  at $y_i$,  $F_i(\bm{\theta}):=\int_{-\infty}^{y_i}p(\tilde{y}_i|\x_i,\bm{\theta})d\tilde{y}_i$.

Let us assume the model $\bm{\theta} \mapsto p_{\bm{\theta}}$ is appropriately smooth and identifiable, the prior puts
positive mass around the true parameter $\bm{\theta}_0$, and the \textit{Fisher information} matrix $\mathcal{I}(\bm{\theta}_0)$ is nondegenerate. Then, the Bernstein-Von Mises (BvM) theorem \citep{hartigan1983asymptotic} asserts that the posterior $\pi_n=:\pi(\bm{\theta}|\mathcal{D}_n)$ converges in \textit{total variation distance} \citep{verdu2014total} to a multivariate normal distribution centered at any estimator $\hat{\bm{\theta}}_n$ satisfying  $\sqrt{n}(\hat{\bm{\theta}}_n-\bm{\theta}_0) \xrightarrow{d} \mathcal{N}\big(0,\mathcal{I}^{-1}(\bm{\theta}_0 )\big)$, 
\begin{equation}
    \parallel \pi_n -\mathcal{N}\big(\hat{\bm{\theta}}_n,n^{-1}\mathcal{I}^{-1}(\bm{\theta}_0)\big)\parallel_{_{TV}}\ \xrightarrow{p_{\bm{\theta}_0}} 0,\quad \text{as}\ n\rightarrow \infty. 
    \label{eq15}
\end{equation}
It is customary to
identify $\hat{\bm{\theta}}_n$ as the maximum likelihood estimate (MLE) in this context. 

Recall that \textit{posterior PITs} may also be written in terms of the predictive CDF as  
 $ p_{\mathrm{post},i}= \int F_i(\bm{\theta}) \pi(\bm{\theta}|\mathcal{D}_n) d\bm{\theta}$. The 
 Taylor expansion of $F_i(\bm{\theta})$ at $\bm{\theta}_0$ to second order yields
\begin{equation}
     F_i(\bm{\theta}) = F_i(\bm{\theta}_0) 
+ \nabla_{\bm{\theta}} F_i(\bm{\theta}_0)^{\top} (\bm{\theta} - \bm{\theta}_0)+ \frac{1}{2}(\bm{\theta} - \bm{\theta}_0)^{\top}
\nabla^2_\theta F_i(\theta_0)(\bm{\theta} - \bm{\theta}_0)+ O(\parallel \theta-\theta_0\parallel^3). 
\label{eq16}
 \end{equation}

Integrating against the full data posterior  $\pi_n:=\pi(\bm{\theta}|\mathcal{D}_n)$ under  BvM conditions gives 
\begin{align}
p_{\mathrm{post},i}&= F_i(\bm{\theta}_0) 
+ \nabla_{\bm{\theta}} F_i(\bm{\theta}_0)^{\top} (\hat{\bm{\theta}}_n - \bm{\theta}_0)+ \frac{1}{2n}tr\big\{\nabla^2_\theta F_i(\bm{\theta}_0)\mathcal{I}^{-1}(\bm{\theta}_0)\big\}+O(n^{-3/2}).
\label{pit}
\end{align}
We refer the reader to Appendix~\ref{sec:appdix_A2} for more details. Applying the same expansion to the LOO posterior $\pi_{n,-i}:=\pi(\bm{\theta}|\mathcal{D}_{n,-i})$, we obtain LOO-PIT values as
\begin{equation}
   p_{\loo,i}= F_i(\bm{\theta}_0) 
+ \nabla_{\bm{\theta}} F_i(\bm{\theta}_0)^{\top} (\hat{\bm{\theta}}_{n,-i} - \bm{\theta}_0)+\frac{1}{2n}tr\big\{\nabla^2_\theta F_i(\bm{\theta}_0)\mathcal{I}^{-1}(\bm{\theta}_0)\big\}+O(n^{-3/2}),
\label{eq17}
\end{equation}
where $\hat{\bm{\theta}}_{n,-i}$ is the MLE using all the observations except $y_i$, $\mathcal{D}_{n,-i}:=\mathcal{D}_{n}\setminus (\x_i,y_i)$. The influence function of M-estimates \citep{ronchetti2009robust}, which we define below, offers a principled analytical mechanism for relating \(\hat{\bm{\theta}}_{n,-i}\) to \(\hat{\bm{\theta}}_{n}\). This relationship arises from approximating the effect of removing a single observation on the MLE.

\begin{definition}[Influence Function]
Let $T$ be a functional that maps a probability distribution $P$ to a parameter value $ \bm{\theta}=T(P)\in \Theta$ (e.g., mean functional $\int x dP(x)$ ). Consider the distributions $ P_{\epsilon,\bm{y}}=(1-\epsilon)P+\epsilon \delta_{\bm{y}}$ where $\delta_{\bm{y}}$ denotes the Dirac distribution in the point $\bm{y}\in  \mathcal{Y}$. Then the influence function ($\IF$) of $T$ at $P$ in the point $y$ is defined as the first order derivative of $T(P_{\epsilon,\bm{y}})$ at $\epsilon=0$,
\begin{equation}
    \IF(\bm{y};T,P)=\lim_{\epsilon\to0} \frac{T(P_{\epsilon,\bm{y}})-T(P)}{\epsilon} .
    \label{eq18}
\end{equation}
\end{definition}
The $\IF$ measures the impact of an infinitesimally small amount of contamination of the original distribution $P$ at the point $\bm{y}$ on the value of $T(P)$. 
Now, let $P_n$ be the empirical distribution of the full data $\mathcal{D}_n$ and $P_{n,-i}$ that of the data $\mathcal{D}_{n,-i}$ which contains all but the $i^{\mathrm{th}}$ observation $\bm{y}_i$. Observing that $P_{n,-i}=\big( 1-(-\frac{1}{n-1})\big)P_n+(-\frac{1}{n-1})\delta_{\bm {y}_i}$ and taking $P_{\epsilon,y}=P_{n,-i},\ \epsilon=-\frac{1}{n-1},$ and  $P=P_n$, Eq.~\eqref{eq18} gives
\begin{equation}
    T(P_{n,-i})-T(P_n)\approx(-\frac{1}{n-1}) \IF(\bm{y}_i;T,P_n) .
     \label{eq19}
\end{equation}
 For an \textit{M-estimate} $\bm{\theta}=T(P)$, the functional $T$ is defined by an implicit equation of the form $\int \psi\big(y;T(P)\big)dP(y)=0$, where $\psi(y;\bm{\theta})=\nabla
_{\theta}\rho(y;\bm{\theta})$ and $\rho(y;\bm{\theta})$ is an arbitrary function \citep[cf. Chap.~3,][]{ronchetti2009robust}. Differentiating the implicit equation using the chain rule and solving for $\dot{T}$, yields the formula for the influence function at $y$, 
\begin{equation}
    \IF(y;T,P)=\frac{\psi\big(y;T(P)\big)}{-\int \nabla_{\theta}\psi\big(y;\theta \big)\big|_{\theta=T(P)} dP(y)}. 
    \label{eq20}
\end{equation}
We are particularly interested in the ordinary MLE, $\hat{\bm{\theta}}_n=T(P_n)$, corresponding to the choice 
\(\psi(y;\bm{\theta}) = \nabla_{\theta} \log p(y \mid \x,\bm{\theta})\) (log score). 

The relation between $\hat{\bm{\theta}}_n$ and $\hat{\bm{\theta}}_{n,-i}$ up to leading order $1/n$ follows from Eq. \eqref{eq19} and Eq.~\eqref{eq20}, 
\begin{equation}
   \hat{\bm{\theta}}_{n,-i}\approx \hat{\bm{\theta}}_n  -\frac{1}{n-1}\IF(y_i), \quad  \IF(y_i)=\mathcal{H}_n^{-1}(\hat{\bm{\theta}}_n) \nabla_{\theta} \log p(y_i|\mathbf{x}_i,\theta)\big|_{\theta=\hat{\bm{\theta}}_n}
\end{equation}
where $\mathcal{H}_n(\theta)=-\mathbb{E}_{P_n}\left[ \nabla^2_{\theta} \log p(y| \x,\theta)\ \big|\ \theta \right]$.
The right-hand side now depends on the full data sample $P_n$. Substituting this expression into Eq.~\eqref{eq17} reveals the underlying structural decomposition of the pointwise LOO-PIT.
\begin{equation}
\begin{split}
p_{\loo,i} &= F_i(\bm{\theta}_0) 
+ \nabla_{\bm{\theta}} F_i(\bm{\theta}_0)^{\top} (\hat{\bm{\theta}}_{n} - \bm{\theta}_0)-\frac{1}{n-1}\nabla_{\bm{\theta}}F_i(\bm{\theta}_0)^{\top}\IF(y_i)\\
& \quad +\frac{1}{2n}tr\big\{\nabla^2_\theta F_i(\bm{\theta}_0)\mathcal{I}^{-1}(\bm{\theta}_0)\big\}+ O(n^{-3/2}).
\end{split}
\label{eq21}
\end{equation}
For notational convenience, we write: \(\bm{f}_i := \nabla_{\bm{\theta}} F_i(\bm{\theta}_0)\ \),
\(\bm{c}_i := \nabla_{\bm{\theta}} F_i(\bm{\theta}_0)^{\top}\IF(y_i)\ \)  and $\bm{Z}_{\theta_0}:=\sqrt{n}(\hat{\bm{\theta}}_n - \bm{\theta}_0)$. Recalling the asymptotic property of the MLE, $\hat{\bm{\theta}}_n$, from BvM theorem, we obtain that, asymptotically,  Eq.~\eqref{pit} and Eq.~\eqref{eq21} reduce to
\begin{equation}
\begin{split}
p_{\post,i}&\approx F_i(\bm{\theta}_0)+\frac{1}{2n}tr\big\{\nabla^2_\theta F_i(\bm{\theta}_0)\mathcal{I}^{-1}(\bm{\theta}_0)\big\}+\frac{1}{\sqrt{n}}\bm{f}_i^{\top}\bm{Z}_{\theta_0}, \\
 p_{\loo,i}&\approx p_{\post,i}-\frac{1}{n-1}c_i, \quad \bm{Z}_{\theta_0}\sim \mathcal{N}\big(0,\mathcal{I}^{-1}(\bm{\theta}_0 )\big).
\end{split}
\label{eq22}
\end{equation}
Thus, conditional on the data, the \textit{posterior PIT}, is the sum of the ideal PIT, $F_i(\bm{\theta}_0)$, and a posterior contribution terms, 
 while LOO-PIT further incorporates an additional self-influence term \(\bm{c}_i\), which is specific to the $i^{\mathrm{th}}$ observation. 
The influence of the double use of the data manifests in the gradient component \(\bm{f}_i := \nabla_{\bm{\theta}} F_i(\bm{\theta}_0)\), which is responsible for jointly pushing the pointwise posterior PIT values toward $1/2$. Conversely, the additional left-out observation-specific term $c_i$, present in the LOO-PIT formulation of Eq.~\eqref{eq22} counteracts this effect, leading to a more uniform spread of LOO-PIT values. 

\paragraph{Marginal uniformity} It follows from Eq.~\eqref{eq22} that, marginally (integrating out the data and noise $\bm{Z}_{\theta_0}$), the \textit{posterior PIT} converges in distribution (at rate $1/\sqrt{n}$) to the ideal PIT, $F_i(\theta_0)$, which is distributed $\Uniform(0,1)$ under a well-specified model. Similarly, as $n\to \infty$ the extra observation-specific term $c_i$ of order $ O(\frac{1}{n-1})$ vanishes, and the LOO-PIT is also asymptotically uniformly distributed.

\subsection{Correlations depend on model complexity and data size}
The dependence among LOO-PIT values arises from the shared posterior mean shift uncertainty $\bm{Z}_{\theta_0}$, which couples all PITs. 
Conditional on $\hat{\bm{\theta}}_n$, each term $c_i$ depends only on $y_i$, so the $\{c_i\}$ are conditionally independent across $i$. Unconditional independence follows from the law of total covariance: the conditional covariance vanishes by independence, and the covariance of the conditional means vanishes because the score equation forces $\mathbb{E}[c_i \mid \hat{\bm{\theta}}_n] = 0$ for all $i$, so that $\mathrm{Cov}(c_i, c_j) = 0$ for $i \neq j$ (see Appendix~\ref{sec:appdix_A3}).
Moreover, since $p_{\post,i}$ depends only weakly on $\bm{y}_i$ through  $F_i(\bm{\theta}_0)$, cross-terms like $\mathrm{Cov}(p_{\post,i},c_j)$ also equal $0$ for $i \neq j$. Thus the leading cross-covariance of LOO-PITs---conditional on $F_i(\bm{\theta}_0)$---is the same order as for the posterior PIT, which follows directly from Eq.~\eqref{eq22},
\begin{equation}
\label{corr}
    \text{Cov}(p_{\loo,i},p_{\loo,j})=\text{Cov}(p_{\post,i},p_{\post,j})= \frac{1}{n}\bm{f}_i^{\top}\mathcal{I}^{-1}(\bm{\theta}_0 )\bm{f}_j+ O(n^{-3/2}),\ \  i\neq j.
\end{equation}
 In matrix form, this is
\begin{equation}
 \text{Cov}(\bm{P}_{\loo})\approx \frac{1}{n}\bm{\mathcal{F}}\mathcal{I}^{-1}(\bm{\theta}_0 )\bm{\mathcal{F}}^{\top}, \quad \bm{\mathcal{F}}=\begin{bmatrix}
\bm{f}_1^{\top} \\ \vdots\\ \bm{f}_n^{\top}
\end{bmatrix}
  \in \mathbb{R}^{n\times p},\ \ \bm{P}_{\loo}=\begin{bmatrix}
p_{\loo,1}\\\vdots\\p_{\loo,n}
\end{bmatrix} \in \mathbb{R}^{n}.
\label{low-rank}
\end{equation}
Eq.~\eqref{corr} shows that pairwise correlations decay at rate $1/n$, yet remain non-negligible for finite sample size $n$. With increasing sample size, the dependencies get weaker, and eventually vanish in the asymptotic limit, leading to uncorrelated PIT values. 

In Eq.~\eqref{low-rank}, the rank of the derived covariance matrix is at most $\min(p,n)$, since $\mathcal{I}^{-1}\in \mathbb{R}^{p\times p}$ and the columns of $\bm{\mathcal{F}}$ live in $\mathbb{R}^{p}$. The dimension $p$ of the parameter space quantifies the model’s complexity--it is the number of directions in which the likelihood can adjust to the data. Under the identifiability condition that the $n$ gradient vectors $\{f_i = \nabla_{\bm{\theta}} F_i(\bm{\theta}_0)\}$ span $\mathbb{R}^p$ which we henceforth assume, $\operatorname{rank}\big\{\bm{\mathcal{F}}\mathcal{I}^{-1}(\bm{\theta}_0 )\bm{\mathcal{F}}^{\top}\big\}=p$ whenever $p < n$.
This means that the induced dependence spans only a (smaller)
$p$-dimensional subspace of the $n$-dimensional PIT space. In other words, 
the PIT values are correlated along a $p$-dimensional plane embedded in the PIT space, and are nearly independent only along the remaining $n-p$ directions orthogonal to this subspace, where changes in the model's parameters cannot explain the data variability. This implies that the model complexity directly determines the extent of LOO-PIT correlations. 

For simple models (i.e., $\frac{p}{n}\rightarrow 0$ ), 
the latter subspace ($p$-dimensional) is very small. Consequently, most of the PIT variation lies in the independent orthogonal space, so that LOO-PIT (and posterior PIT) values appear distributed almost  $i.i.d.\ \Uniform(0,1)$. 
For complex models (i.e., $p$ not small relative to the data size $n$ or $p>n$), the $p$-dimensional subspace expands, that is, the variation induced by the posterior mean shift term projects into most directions of the PIT space, resulting in stronger and more structured dependence among the LOO-PIT (and posterior PIT) values. 

In Bayesian modelling, model complexity is usually not directly determined by the raw number of model parameters $p$, but rather the \textit{effective number of parameters} in the model, which can be estimated using  \(\hat{p}_{\loo}\) as defined by \citet{vehtari2017practical}. Using this, the above results can be summarized in two points. (i) For fixed model complexity, \( \hat{p}_{\loo}\), the LOO-PIT dependence vanishes asymptotically as \(n \to \infty\). (ii) For fixed sample size \(n\), the dependence becomes negligible when the ratio \(\hat{p}_{\loo}/n\) approaches zero. Moreover, when this ratio is already sufficiently low, the posterior PIT and the LOO-PIT agree closely and can be used interchangeably. 

\subsection{Under model misspecification}
Previously, PIT values were studied under the assumption of correctly specified model. 
We now examine, their asymptotic behavior under misspecification, where 
the true data-generating distribution $P_0$ does not necessarily belong to the 
parametric family $\{p_{\bm{\theta}} : \bm{\theta} \in \Theta\}$. In this situation, 
\cite{kleijn2012bernstein} show that the Bernstein-Von Mises theorem of 
Eq.~\eqref{eq15} can be extended in the form,
\begin{equation}
       \parallel \pi_n -\mathcal{N}\big(\hat{\bm{\theta}}_n,n^{-1}V(\bm{\theta}^*)\big)\parallel_{_{TV}}\ \xrightarrow{p_{\bm{\theta}_0}} 0,\quad \ n\rightarrow \infty,
    \label{ubvm}
\end{equation}
where $\bm{\theta}^*$ is the parameter value closest to the truth in Kullback-Leibler divergence, $\bm{\theta}^*=\arg\min_{\theta\in \Theta} \operatorname{KL}\{P_0||p_{\bm{\theta}}\}$ (provided it exists and is unique ), and the asymptotic covariance is given by the \textit{sandwich formula} $V(\bm{\theta}^*)=J^{-1}(\bm{\theta}^*)D(\bm{\theta}^*)J^{-1}(\bm{\theta}^*)$ with $J(\bm{\theta})=-\mathbb{E}_{P_0}[\nabla_{\bm{\theta}}^2\log p_{\bm{\theta}} ]$ the generalized \textit{Fisher information} and $D(\bm{\theta})=\mathbb{E}_{P_0}[\nabla_{\bm{\theta}}\log p_{\bm{\theta}} (\nabla_{\bm{\theta}}\log p_{\bm{\theta}})^{\top}].$
Eq.~\eqref{eq21} then applies analogously, but now $\bm{\theta}^*$ replaces $ \bm{\theta}_0,$ and the mean shift uncertainty is rescaled according to the  \textit{sandwich covariance}, $Z_{\theta^*}:=\sqrt{n}(\hat{\bm{\theta}}_n - \bm{\theta}^*) \sim 
\mathcal{N}\big(0,\,
V(\bm{\theta}^*)
\big)$: 
\begin{equation*}
    p_{\loo,i}\approx F_i(\bm{\theta}^*)  +\frac{1}{2n}tr\big\{\nabla^2_\theta F_i(\bm{\theta}^*)V(\bm{\theta}^*)\big\}+ \frac{1}{\sqrt{n}}\nabla_{\bm{\theta}} F_i(\bm{\theta}^*)^{\top}Z_{\bm{\theta}^*}-\frac{1}{n-1}\nabla_{\bm{\theta}}F_i(\bm{\theta}^*)^{\top}\IF(y_i)
   .
\end{equation*}
The low-rank dependence is inflated relative to the well-specified case when $V\neq J$. More importantly, in the limit, $n\rightarrow \infty$, both LOO-PIT and posterior PIT converge  in distribution at rate $1/\sqrt{n}$ to $F_i(\bm{\theta}^*),$ which in general need not be uniformly distributed on $(0,1)$, thereby revealing the lack of fit. Next, we discuss pointwise testing procedures that can account for the type of dependency present among LOO-PIT values.

\section{Dependence-aware uniformity tests }
\label{sec:dependence-aware}
In this section, we introduce three uniformity testing procedures that can account for possible dependence among the sample values  $u_1,\dots,u_n \in (0,1)$: \textit{Pointwise Order Tests Combination} (POT-C), \textit{Pointwise Rank-based Individual Tests Combination} (PRIT-C), and \textit{Pointwise Inverse-CDF Evaluation Tests Combination} (PIET-C). 
Like the graphical approaches discussed above, POT-C and PRIT-C use the beta and binomial distributions, respectively, to compute pointwise order-specific and rank-specific test \(p\)-values $p^{(i)}$ for $i=1,\cdots,n$. The individual \(p\)-values are then combined into a single global $p$-value in a way that explicitly accounts for the induced dependence among them. On the other hand, the PIET-C procedure combines pointwise $p$-values from tests on pseudo-values obtained by transforming the $u_i's$ through the \textit{inverse CDF} of a continuous variable $X,$ with the test's distribution tied to that variable. In the sequel, we assume that the values $u_1,\dots,u_n$ being tested for uniformity are potentially correlated under the null. 
We first describe how pointwise tests are conducted for each approach in Section~\ref{sec:pointwise tests}, and then describe how the individual tests are combined to perform an overall test in Section~\ref{sec:combining tests}. 

\subsection{Pointwise testing}
\label{sec:pointwise tests}

\paragraph{POT-C} Individual tests are based on beta distributions. The $i^{\mathrm{th} }$ order statistic $u_{_{(i)}}$---($u_{_{(1)}}\leq u_{_{(2)}}\dots\leq u_{_{(n)}}$)---of the statistical continuous valued sample $u_1,\dots,u_n$ follows a beta distribution under uniformity. This provides a natural way for constructing marginal uniformity tests by comparing the observed value of \( u_{_{(i)}}\) against its theoretical null distribution \( \Beta(i,n+1-i) \) for each \( i \). 
If the observed order statistic falls in the tails of this distribution, the corresponding \textit{$p$-value} \( p^{(i)} \) provides evidence against uniformity at that order. Denote the value of the cumulative beta distribution function with parameters $\alpha=i$ and $\beta=n+1-i$ at $u\in(0,1)$ by $\Beta(u\mid i,n+1-i)$. The $p$-value is obtained by first identifying in which tail the observed statistic falls and how extreme it is in that tail. This probability is then doubled to account for the total probability of observing such an extreme value in either direction under the null : 
\begin{equation}
    p^{(i)}=2\times \min\big\{ 1-\Beta(u_{_{(i)}}\mid i,n+1-i),\Beta(u_{_{(i)}}\mid i,n+1-i)\big\},\quad i=1\ldots,n. 
    \label{pbeta}
\end{equation}

\paragraph{PRIT-C} Individual tests are based on binomial distributions. For a given sample $u_1,\ldots,u_n$, the ECDF evaluated at a point $z_i \in (0,1)$ is $F(z_i) = \frac{1}{n}\sum_{j=1}^n \mathbb{I}(u_j \le z_i)$. 
If $u_j$ are discrete (e.g., rank-based PIT values), choosing the partition points to form a subset of these category values yields $\Pr(u_j \le z_i) = z_i$ for all $j$ \citep{sailynoja2022graphical}. It then follows that the scaled ECDF values are binomially distributed as $nF(z_i) \sim \mathrm{Bin}(n, z_i)$. This last fact allows us to construct individual $p$-values  for ranks similarly to before, 
\begin{equation}
    \label{prit-c}
       p^{(i)}=2\times \min\big\{ 1-\Bin(nF(z_i)-1\mid n,z_i),\Bin(nF(z_i)\mid n,z_i)\big\},\quad i=1\ldots,n. 
\end{equation}
\paragraph{PIET-C} Individual tests are constructed with respect to any continuous distribution.
We assume that $u_i$ are continuous and denote by $F^{-1}_X$ the inverse CDF of a continuous random variable $X$. The pseudo-values
obtained via the transformation $F^{-1}_X(u_i)$ can then be regarded as samples
from the distribution of $X$ under the null hypothesis. This follows from applying
the PIT (Section~\ref{sec:pit}) in reverse: if $u_1,\ldots,u_n$ are marginally
uniform, then $F^{-1}_X(u_1),\ldots,F^{-1}_X(u_n)$ are marginally distributed as $X$. We leverage this principle to perform pointwise uniformity checks by comparing
each pseudo-value $F^{-1}_X(u_i)$---in the X's space---to its theoretical distribution $F_X$. \textit{p}-values $p^{(i)}$ are computed
as in Eq.~\eqref{pbeta},
\begin{equation}
    p^{(i)}=2\times \min\big\{\Pr\big(X>F_X^{-1}(u_i)\big),1-\Pr\big(X>F_X^{-1}(u_i)\big)\big\},\quad i=1\ldots,n.
\end{equation}

Although valid (marginal) tests will produce uniformly distributed $p$-values, $p^{(i)}\sim \Uniform(0,1)$ under the null, they will inherit the dependence induced by the type of dependence structure present among $u_1,\ldots,u_n$, with additional dependency that arises from the ordering (or ranking) transformation in the case of order- and rank-based tests. 

The rationale for pointwise uniformity checks, rather than a joint uniformity test, 
is to avoid specifying the full null joint distribution of $(u_1,\ldots,u_n)$, which 
is intractable under dependence. Each individual test is marginally valid (approximately) by construction. The next step is to aggregate individual $p$-values, into a single  global $p$-value,$\ p^*$, that will serve as the overall evidence against the global null hypothesis.

\subsection{Combining individual tests}
\label{sec:combining tests}
To combine multiple $p$-values, $p^{(1)}, \ldots, p^{(n)}$, 
 \cite{tippett1931methods} proposed using the minimal \(p\)-value method, which computes a global \(p\)-value based on the minimum order statistic \(p_{_{(1)}} = \min\{p^{(1)}, \ldots, p^{(n)}\}\). Although this method is simple and computationally attractive, it becomes relatively conservative in terms of test power when applied to correlated test \(p\)-values. This is because most of the useful information from the remaining p-values are disregarded. A natural way to achieve higher test power would be to combine information from all \(p\)-values, rather than focusing solely on the smallest one.

We combine the $p$-values from above  using
the Cauchy combination test (CCT) \citep{liu2020cauchy}---a test statistic defined as a weighted sum of transformed $p$-values, where the $p$-values are transformed to follow a standard Cauchy distribution. In particular, equal weighting is employed here, resulting in an average: 
\begin{equation}
    T=\frac{1}{n}\sum_{i=1}^{n}\tan\big\{ (0.5-p^{(i)})\pi \big\}.
    \label{cct}
\end{equation} 
Given $p^{(i)}\sim \Uniform(0,1)$, the component $\tan\big\{ (0.5-p^{(i)})\pi\}$ follows a standard Cauchy distribution that we denote by $C_0$. With arbitrary correlations between $p^{(i)}$’s, the test statistic $T$ is defined as a weighted sum of  correlated standard Cauchy variables and may not have a closed form null distribution. \cite{liu2020cauchy} argued that the correlations only have limited impact on the tail of $T$ because of the heaviness of the Cauchy tail, and proved that the tail of the null distribution of the test statistic $T$ is approximately Cauchy under arbitrary correlation structures: $  \lim_{\alpha\to 0} \frac{P\{T>t_{\alpha}\}}{P\{C_0>t_{\alpha}\}}=1$, where $t_{\alpha}$ is the upper $\alpha$-quantile of $C_0$.
This suggests that for moderately small significance level $\alpha$, we can use the standard Cauchy CDF to approximate the $p$-value of the test based on $T$ as, 
\begin{equation}
    p^*=1-F_\mathrm{Cauchy}\left(T\right).
    \label{p-cct}
\end{equation}
The aggregated $p$-value $p^*$ now stands as the evidence against the global null hypothesis. It is then compared to a prespecified level $\alpha$ to decide whether to reject the null. Although $p^*$ is not uniform under the null, it follows from the tail convergence that for
practically relevant values of $\alpha$, e.g. $\alpha\in (0,0.05)$, the Type I error is near the target level, $\mathbb{P}(p^*\leq \alpha)\approx \alpha $. \cite{chen2025truncated} proposed the \textit{truncated Cauchy combination test} (TCCT), $T_{_{TCCT}}=\frac{1}{n}\sum_{i=1}^{n}\tan\big\{ (0.5-p^{(i)})\pi \big\}\mathbb{I}(p^{(i)}<0.5)$, as an improvement to the original Cauchy test and similarly demonstrated that its tail probability has the desired asymptotic property of Cauchy tail probability so that under the null, the $p$-value can again be approximated as in Eq. \eqref{p-cct}.
In our experiments, we find that TCCT always has slightly improved test power compared to the original Cauchy combination method. 

\subsection{Graphical representation}
\label{sec:graphical_rep}
To complement the developed testing procedures, we present an intuitive graphical representation for visualizing pointwise test results. It involves an automated procedure that emphasizes specific parts of the \textit{ECDF} plot through color coding. The goal is to provide users with visual cues that communicate where local discrepancies are prominent. 

Since the Cauchy test statistic of Eq.~\eqref{cct} is the average of the Cauchy transformed p-values $t_i:=\tan\big\{ (0.5-p^{(i)})\pi \big\}$, the idea is to quantify the relative impact of each individual test $i$ on the overall test by calculating how influential the corresponding value $t_i$ in the Cauchy space is for that average and depict that influence through color.
The Shapley values \citep{shapley1953value} defined below provide a principled approach for quantifying these point-specific influences in a way that reflects local miscalibration. 
\begin{definition}[Shapley Value]
Consider an $n$-person game $(N,v)$, where $N = \{1,\dots,n\}$ is a finite set of players and 
$v : 2^N \to \mathbb{R}$ is a characteristic function that assigns a value to each coalition $S \subseteq N$.
For any player $i \in N$, the \emph{Shapley value}, $\phi_i(v)$, of player $i$ is defined as the expected marginal contribution of player $i$ when the players join the coalition in a uniformly random order:
\begin{equation}
    \phi_i(v)
= \sum_{S \subseteq N \setminus \{i\}}
\frac{|S|!\,(n - |S| - 1)!}{n!}
\big( v(S \cup \{i\}) - v(S) \big),
\label{shapley}
\end{equation}
\end{definition}

The Shapley value is a solution concept from cooperative game theory that assigns to each player $i$ a fair share of the total value created by all players. Based on this, one can turn the question of quantifying individual influences of each test $i$ to dividing the value of the average of the Cauchy values---observed Cauchy statistic $T$ in Eq.~\eqref{cct}---amongst them in a cooperative game framework with the following set up: 
\begin{itemize}
    \item The player set is \(N = \{1,\ldots,n\}\), where each index \(i\) corresponds to a point \(u_i\) and is also associated with the corresponding tested quantity—namely, its order \(u_{(\sigma(i))}\) for POT-C, its rank \(nF(u_i)\) for PRIT-C, or its transformation \(F_X(u_i)\) for PIET-C.

    \item  The value function is defined as, $\quad v(S) =
\begin{cases}
\displaystyle \frac{1}{|S|}\sum_{S} t_i, & S \neq \varnothing \\
0, & S = \varnothing,
\end{cases}$
\end{itemize}
where $|S|$ is the cardinality of the coalition set $S\subseteq N$, $t_i:=\tan\big\{ (0.5-p^{(i)})\pi \big\}$, and $p^{(i)}$ is the pointwise $p$-value associated with $u_i$ computed in any of the form discussed in Section~\ref{sec:pointwise tests}. Under this game set up, the expected marginal contribution of each $u_i$ as formulated in Eq.~\eqref{shapley} can be calculated deterministically:
\begin{equation}
    \phi_i(v)=\frac{1}{n}t_i + \frac{H_n - 1}{n}\Big(t_i - \frac{1}{n-1}\sum_{j \neq i} t_j\Big),
    \label{shap}
\end{equation}
where $H_n = \sum_{k=1}^n \frac{1}{k}$ is the harmonic number (see Appendix~\ref{sec:appdix_A4} ). The $\phi_i(v)$'s then tell us how influential each individual test $i$ (or point $u_i$) is to the value of the grand coalition, that is, $v(N)=\frac{1}{n}\sum_{i=1}^{n}\tan\big\{ (0.5-p^{(i)})\pi \big\}$---the Cauchy test statistic. The total value created by the grand coalition is fully distributed among all $u_i$, $\sum_{i=1}^{n}\phi_i(v)=v(N)$. Moreover, any two points that contribute identically to every coalition receive identical Shapley values---reflecting equal influence. These properties follow directly from the efficiency and symmetry axioms satisfied by the Shapley's solution \citep{shapley1953value}.

Because $v(\emptyset)=0$ (coalition of no players) and $v$ is not a monotonic function (i.e., adding players can reduce the coalition’s value), the marginal contributions, $\phi_i(v)$, include negative terms.  All the points $u_i$ for which $\phi_i(v)$ is non-negative are a priori suspicious. This is because they contribute to pushing $v(N)$ deep into the upper tail of the standard Cauchy distribution. When the CCT rejects the null, this likely indicates one of two scenarios. Either a very small number of points have very large positive influences, as measured by \(\phi_i(v)\), which completely dominate the average and push it deep into the Cauchy tail; or, due to dependence among the \(u_{i}\), many points are close in the Cauchy space and thus exhibit similar positive influences, and their combined effect jointly drives the average far into the tail. The highlighting strategy is as follows: 

\begin{figure}[tp]
    \centering    \includegraphics[width=\linewidth]{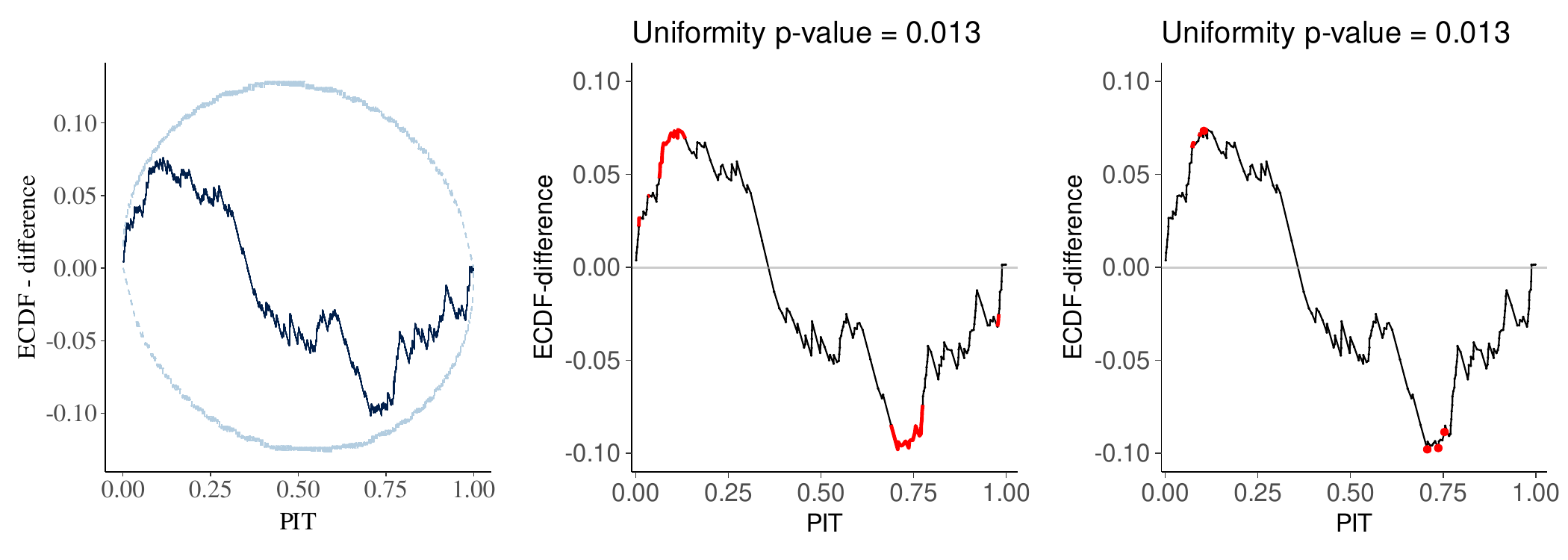}
    \caption{Three visualizations depicting the same sample of $200$ PIT values. To asses uniformity of the sample, (left) the ECDF difference plot with 95\%
coverage for the ECDF  shows the sample staying within the given limits. Meanwhile, the proposed color coding based graphical procedure detected issues with the sample and triggered identification of statistically suspicious parts of the ECDF (in red): (middle) $\gamma=0,\ \text{(right)}\ \gamma=\frac{\max\phi_i(v)}{2}$.}
    \label{fig:6}
\end{figure}

\textbf{Influential regions of the ECDF plot}. Given the point-specific contributions as measured by the Shapley values \(\phi_i(v)\), we introduce a free parameter $\gamma\in [0, \max_{_i}\phi_i(v)]$, which determines the level of influence we care about. The influential regions relative to $\gamma$ that we denote by $\mathcal{IR}(\gamma)$, are those parts of the ECDF plot determined by points $u_{i}$ whose expected marginal contribution  exceeds $\gamma$: 
$$ \mathcal{IR}(\gamma)=\{u_{i}\ |\ \phi_i(v)>\gamma \}. $$
Those points are then automatically, identified and highlighted in the ECDF plot. When $\gamma=0$, all suspicious points will be highlighted. Increasing $\gamma$ beyond $0$ emphasizes only the regions that most contribute to the overall test.  We recommend setting $\gamma=0$ when $p*<\alpha$ to enable a more comprehensive visualization of local departures from uniformity. This procedure is computationally very efficient compared to, for example, ECDF-envelope test, which requires coverage adjustment through simulation or optimization to account for multiplicity when forming the $1-\alpha$ confidence envelope for the ECDF. Figure~\ref{fig:6} illustrates the visual approach using the POT-C test for a random example with AR-type (band-diagonal covariance) dependence in the PIT values.

\subsection{Some practical considerations for use with LOO-PIT}
\label{sec:properties}
We discuss some considerations relevant to the application of the proposed methods to LOO-PIT in terms of model criticism, with attention to the nature of the PIT values. 

By construction, the PIET-C and POT-C approaches require that pointwise PIT values be continuous---calculated through CDF evaluation---for stable uniformity checking. This poses no difficulty, as the data model is typically tractable and its distribution CDF, $F(y_i|\x,\theta):=\int_{-\infty}^{y_i} p( \tilde{y}|\x, \theta) d\tilde{y}$, 
can be easily evaluated given posterior draws. For discrete data models, these can be random points picked uniformly between the CDF jump at observations $y_i$ to recover the $\Uniform(0,1)$ characterization of proper PIT values \citep{Czado-Gneiting-Held:2009}: $p_i=F(y_i-1|\x,\theta)+v\left(F(y_i|\x,\theta)-F(y_i-1|\x,\theta)\right),\ \text{with}\ v\sim \Uniform(0,1)$.
Using discrete PITs (calculated as normalized ranks Eq.~\ref{eq12}) with continuous uniformity test would slightly affect test power and Type I error control. This is because the continuous version is expected to provide more accuracy for values near $0$ and $1$, whereas discrete values inevitably sacrifice some accuracy there. However, continuous PITs offer no advantage for PRIT-C---rank-based PITs should be preferred with this test, as it is specifically designed for  discrete ranks. 

When using PSIS-based LOO computation \citep{vehtari2017practical}, LOO-PIT values are computed as weighted averages of PITs across posterior draws using the continuous PSIS-LOO weights. Even starting from discrete (rank-based ) PITs, this weighting yields LOO-PIT values that are near-continuous violating assumption in PRIT-C. However, if rank-based PIT value is 0 or 1, weighting does not change the value, and these 0 and 1 values violate assumption in POT-C and PIET-C. PIT values 0 and 1 can be avoided by computing PITs with parametric family CDFs (and randomization in case of discrete distributions) or by combining the rank based approach with extreme value analysis based PIT values for the extreme ranks.

\paragraph{Remark.}
We note that under strong dependence, the individual $p$-values of POT-C and PRIT-C may not exactly 
follow $\Uniform(0,1)$ under the null. Dependence reduces the dispersion of order 
statistics and ECDF counts relative to the i.i.d. setting, so the beta and binomial 
reference distributions are no longer exactly calibrated, rendering individual $p$-values that are stochastically smaller than $\Uniform(0,1)$. As a result, both procedures provide valid but  conservative Type~I error control in presence of strong correlation (see Table~\ref{tab1}). In the context of model assessment, where guarding against unwarranted model rejection is often desirable, this mild conservativeness is not detrimental, especially since both tests still retain competitive power to detect non-uniformity of LOO-PIT values. 

\section{Related methods}
\label{sec:other-methods}
As our primary focus is on providing a principled testing procedure that properly accounts for potential dependency in PIT values, we want to also ensure that the overall performance of our tests is, if not the best,
competitive with closely related approaches---specifically those that, like LOO-PIT, condition on nearly all the data, unlike methods such as SPC or HPC, which leave a substantial  portion of the data out.

\subsection{Sampled Posterior p-value (SPP) }
The sampled posterior $p$-value (SPP) method developed in \cite{johnson2007bayesian} calculates a $p$-value using a single parameter vector, $\bm{\theta}^*\sim p(\bm{\theta}|\bm{y})$, simulated from the posterior. Although this simplification is computationally efficient, the resulting $p$-value is inherently random as it can very depend upon the posterior draw $\bm{\theta}^*$. However, several studies \citep{zhang2014comparative,gosselin2011new} argued that it is exactly this random sampling of parameters that removes the influence of double use of the data and thus may be a preferable choice over PPC for model criticism. In fact, sampled posterior $p$-values are guaranteed to at least have an asymptotic uniform distribution given the null model \citep{gosselin2011new}. Focusing on pointwise predictive checks, the method implies computing pointwise PIT values using a single draw \( \bm{\theta}^* \sim p(\bm{\theta} | \bm{y}) \) from the posterior, leading to the following formulation:
\begin{equation}
\label{spp}
    u_i=\int_{-\infty}^{y_i} 
p( \tilde{y}_i|\x_i,\theta^*) d\tilde{y}_i,\quad i=1,\ldots,n ,
\end{equation}
where \( p(y |\x, \theta) \) is the observation model, which usually admits a closed-form expression in practice, making the integral straightforward to evaluate. The sampled posterior $p$-value ( corresponding to draw $\theta^*$) is the $p$-value computed from testing $\bm{U}=(u_1,\ldots,u_n)$.

\subsection{Uniform Parameterization  Checks (UPCs)}
 Uniform parameterization checks (UPCs), though presented in a more general framework by \citet{covington2025bayesian}, amount to aggregating SPP results across multiple posterior draws using the Cauchy combination method to account for the data-induced dependence between the tests. Aggregation should reduce randomness and thus increase power to detect model misspecification.

UPC aggregates dependent uniformity test $p$-values computed with different posterior draws. LOO-PIT integrates over the posterior and our procedures can be used to make single dependent uniformity test. Both approaches can be assumed to reduce randomness compared to SPP, and the difference is in the order of aggregation (integration) and testing.

\section{Empirical Study}
\label{sec:experiments}
We study the proposed testing procedures on a variety of numerical experiments.
On synthetic data, we first investigate the Type I error rates control under the null hypothesis of a correctly specified generalized additive mixed-effects model (GAMM). By independently increasing model flexibility and sample size, we create situations in which the LOO-PIT dependence becomes either stronger or weaker, so as to systematically examine its impact on tests calibration. Next, in simulation studies, we assess the power of the POT-C, PRIT-C, and PIET-C procedures for detecting model miscalibration under scenarios where the model's LOO-PIT values are likely to be strongly correlated. The considered examples include hierarchical models with  large effective number of parameters relative to the number of observations.
The study also includes a comprehensive comparison of our methods with the SPP and UPC approaches. In addition, we consider the Anderson–Darling (AD) and Kolmogorov–Smirnov (KS) tests as examples of standard uniformity checks assuming independent pointwise LOO-PIT values. 
 
All simulations were done in R \citep{r2020r} using the brms package \citep{burkner2017brms} together
with the probabilistic programming language Stan \citep{team2020rstan, carpenter2017stan} for posterior inference, and the
loo R package \citep{vehtari2015loo} for the PSIS-LOO computations. 
Posterior sampling was performed with default  Hamiltonian Monte Carlo (HMC) \citep{hoffman2014no,team2020rstan} settings in brms. UPC was implemented using all available posterior draws in the considered examples to ensure maximum power and a fair comparison. 

 Code to reproduce results is available at \href{https://github.com/hermanFTT/LOO-PIT-Project}{https://github.com/hermanFTT/LOO-PIT}. For illustration of LOO-PIT-ECDF plots with POT-C and Shapley values in real data analysis case studies, see \citet{Bayesian-Workflow_2026}  (Ch. 17, 18, and 24, with code available at \url{https://avehtari.github.io/Bayesian-Workflow/}).

\subsection{Type I error control: effects of model complexity and sample size}

\label{sec:type_1_error_control}
In this section, we assess how the LOO-PIT dependence, influenced by model complexity and sample size (refer to Section~\ref{sec:loo-pit dependence}), affects Type I error under the null model.
 Synthetic datasets are generated from a Gaussian hierarchical data-generating process with group-level heterogeneity and a smooth nonlinear effect of a continuous covariate. Specifically, observations arise from, $y_{ig} \mid x_{ig}, \mu_g \sim \mathcal{N}( f(x_{ig}) + \mu_g, \sigma^2)$. 
 $\mu_g\sim \mathcal{N}(0,0.2^2)$ is group-specific effect, 
 $x \sim \mathcal{U}(-2,2)$, and $f(x)=0.6\sin(\pi x)-0.3x^2+0.1x^3$ defines a smooth nonlinear effect combining oscillatory and polynomial components. For each dataset, $G=50$ groups are observed with $m$ observations per group. We then posit a spline-based mixed-effects model with independent priors of the form:
 \begin{align*}
y_{ig} 
&\sim \mathcal{N}(s(x_{ig}) + \mu_g, \sigma^2) \\
\mu_g  
&\sim \mathcal{N}(\mu_0, \tau^2)\\
\mu_0&\sim \mathcal{N}(0, 1),\ \ \ \lambda \sim \text{Exp}(5),\ \ \sigma,\tau 
\sim \text{Exp}(1)
\end{align*}
for all $i,g$, where $\lambda$ is the smoothing (regularization) parameter for the spline component. This model was carefully selected to span a wide range of effective complexities, while already providing excellent agreement with the true data-generating process for small values of the spline basis dimension ($k\leq5$). 

To isolate the role of dependence among LOO-PIT values, we consider two complementary experimental regimes. In the first, we fix the number, $m$,  of observations per group ($m=10$) and increase model flexibility by varying the spline basis dimension $k$, which leads to increasing and stronger correlations among LOO-PIT values. In the second, we fix the effective model complexity ($k=12$) and increase the number $m$ of observations per group, which is expected to weaken the LOO-PIT dependence. For each configuration, we simulate $S=5000$ independent datasets from the true DGP, fit the posited model, compute the LOO-PIT values, and check for their uniformity using the proposed tests. The Type I error rate is then empirically estimated as the proportion of rejections across the simulations at level $\alpha=0.05$. A calibrated test would have a rejection rate that closely matches the nominal level $\alpha$. 
\begin{table}[tp]
\centering
\small
\begin{tabular}{cccccccccc}
$k$ & $m$ & PIET-C & POT-C & PRIT-C &SPP& UPC+KS& UPC+AD& KS & AD \\
\midrule
$48$ & $10$  &0.048 & 0.006 &0.001 & 0.056 &0.066  & 0.092 &0.002  &0.004 \\
$24$ & $10$ &0.051  & 0.022 & 0.011 & 0.060 & 0.076 & 0.094 &0.000 &0.019\\
$6$ &$10$  & 0.050 &  0.039& 0.022 & 0.052 & 0.083 & 0.093&0.000 &0.032\\
\midrule
$12$ & $5$ & 0.050 & 0.004 & 0.002 & 0.048 & 0.063 &0.090  &0.000 &0.000\\
$12$ & $10$ & 0.047 & 0.033 & 0.013 & 0.045 & 0.083 &0.082  & 0.000&0.030\\
$12$ &$15$  &0.052  & 0.068 & 0.042 & 0.044 & 0.088 & 0.107 &0.001 &0.056\\
\end{tabular}
\caption{Type I error control (target is 0.05) for the different tests, varying model complexity with \#spline basis function $k$ and  sample size with \#observations per group $m$. Results are based on S=5000 simulation trials for each combination.}
\label{tab1}
\end{table}

\begin{figure}[tp]
    \centering    \includegraphics[width=\linewidth]{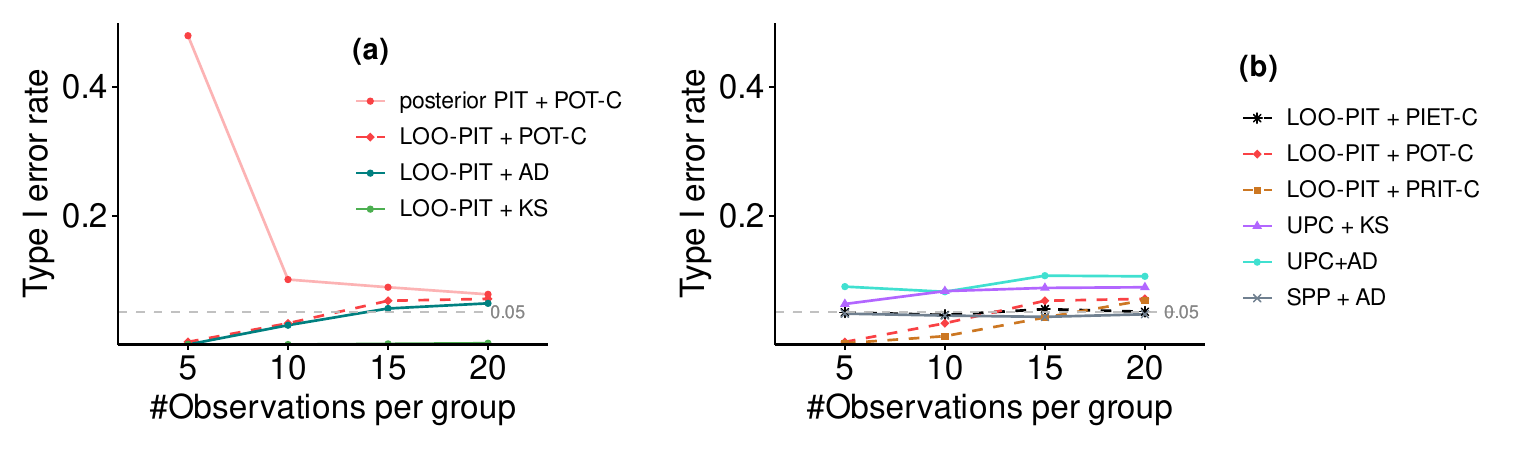}
    \caption{Visualizing Type I error performance for increasing \#observations per group $m$. $k$ was fixed to $12$. \textbf{(a)} Posterior PIT (solid red line) shows increasing concordance with LOO-PIT as sample size grows larger.} 
    \label{fig:1}
\end{figure}

Table~\ref{tab1} and Fig.~\ref{fig:1} provide a comparative summary of the results. Overall, PIET-C consistently maintains good Type I error control, remaining well calibrated across all levels of dependence among the LOO-PIT values. In contrast, POT-C and PRIT-C often tend to be somewhat conservative under the null as suspected: their empirical Type I error rates fall noticeably below the nominal level \(\alpha = 0.05\) when model flexibility is very high relative to sample size, that is, when correlations among LOO-PIT values are substantial. Nevertheless, this effect fades out as model flexibility decreases or the sample size increases, and the Type I error rates gradually approach the nominal level. A similar trend is observed for the AD test, whereas the KS test becomes overly conservative in the presence of dependence, with Type I error rates collapsing to near zero values. In comparison, when AD or KS tests are applied within the UPC methodology  prior to aggregation, both approaches yield inflated Type I error rates across all settings. Interestingly, we noticed in the experiments that this inflation is more pronounced for the AD-based UPC procedure as the sample size increases, while the KS-based UPC remains more stable, with the  inflated Type I error bounded at $0.1$ in the worst case. Besides, in accordance with the theory that posterior PIT and LOO-PIT values may be used interchangeably for simple models (see Section~\ref{sec:loo-pit dependence}), we observe increasing agreement between test outcomes based on posterior PIT and LOO-PIT values as the sample size, \(n = m \times G\), grows ( i.e, $m$ increases),  eventually coinciding when the ratio $\frac{\hat{p}_{\loo}}{n}$ is sufficiently low as shown in Figure~\ref{fig:1}~(a).  

\subsection{Power performance assessment under high model complexity}
We use simulations studies to examine the power of our procedures by calculating the percentage of times our testing procedures rejects the hypothesized model given that the simulated data is generated according to the alternative distributions. 
Attention is restricted to situations where we know the model is so flexible that the LOO-PIT values will exhibit strong correlations, corresponding to settings in which the ratio $\frac{\hat{p}_{\loo}}{n}$ is far from $0$.  
We experimented also with models getting less complex (i.e, $\frac{\hat{p}_{\loo}}{n}\rightarrow0$ ) and the general pattern of the results holds throughout, with our procedures having even higher power and more accurate Type I error control compared to the other methods studied. 

The models considered are hierarchical models for continuous and count data, unless otherwise stated, with \(G = 50\ \text{groups},\  g\in \{1,\cdots,G\}\),  and \(m = 5\) observations per group,  yielding a setting in which the effective model complexity $\hat{p}_{\loo}$ is not small relative to the data size.  We use controlled alternative data distributions that deviate progressively from the assumed model, so we can systematically evaluate the power of predictive checks. Power analysis is based on \(S = 5000\) simulations, and \(\alpha = 0.05\). 

\paragraph{A simple hierarchical model for continuous data}
We return to the example model structure  in Section~\ref{sec:type_1_error_control}, but now dropping the nonlinear smooth term, resulting in a simple normal random effects model with weakly informative priors. 
\begin{align}
    y_{ig}&\sim \mathcal{N}(\mu_{g},\sigma^2) \label{shm},\\
    \mu_g&\sim\mathcal{N}(\mu_0,\tau^2), \nonumber\\
    \mu_0&\sim \mathcal{N}(0,1) ,\nonumber \ \  \sigma,\tau\sim \text{Exp}(1) \nonumber
\end{align}
for all $i,g$. We will consider various choices for the true data generating process (DGP). For each scenario, we simulate $S=5000$ datasets from the true DGP, and for each dataset, we fit the above hypothesized normal model and look at the rejection rates across simulations based on the LOO-PIT diagnostic and competing methods considered. 
\begin{itemize}[topsep=0pt,partopsep=2pt,itemsep=2pt,parsep=2pt]
\item\textbf{Scenario \#1: Heavy-tailed}. %
    We begin with the case where true DGP are \textit{Student's t} distributions, $t_{\nu}$, as a controlled form of misspecification. By varying the degree of freedom $\nu$, we smoothly interpolate between distributions close to Normal (large $\nu$) and increasingly heavy-tailed distributions (small $\nu$). Data are generated as, $y_{ig}|\tau,\sigma, \mu_0\sim t_\nu(\mu_g,\sigma^2)$. with the population level variance, $\tau=1.96$  and the within group variance $\sigma=0.06$ corresponding to the 5th and 95th percentile of the truncated standard normal priors. The population mean was set to $\mu_0=0$.  
\item \textbf{Scenario \#2: Skewed}. 
    Observations are log-normally distributed as, \(y_{ig} \mid \tau,\mu_0,\, \sigma \sim \mathrm{LogNorm}(\mu_g, \sigma^2)\), with the population mean \(\mu_0 = 1\), the group-level variance \(\tau = 0.96\), and the observation noise \(\sigma\) is defined on the log scale. To control for the intensity of skewness, we progressively decrease the within-group variance \(\sigma\) toward \(0\). In the limit as \(\sigma \rightarrow 0\), the induced skewness vanishes. 
\item \textbf{Scenario \#3: Light-tailed}.
    Data are generated according to a \textit{symmetric generalized normal} distribution of the form,  \(y_{ig} \mid \tau,\mu_0,\ \beta \sim \mathrm{GN}(\mu_g,\alpha,\beta)\), where $\alpha=0.31$ is the scaling parameter and $\beta$ is the shape controlling the tail behavior and peakedness of the distribution. For $\beta =2$ the generalized normal reduces exactly to the normal distribution, and with increasing $\beta$ beyond $2$, it becomes light-tailed compared to normal. $\mu_0\ \text{and}\ \tau$ were set to $0$ and $1.96$ respectively. 
\end{itemize}

\begin{figure}[tp]
    \centering    
    \includegraphics[width=\linewidth]{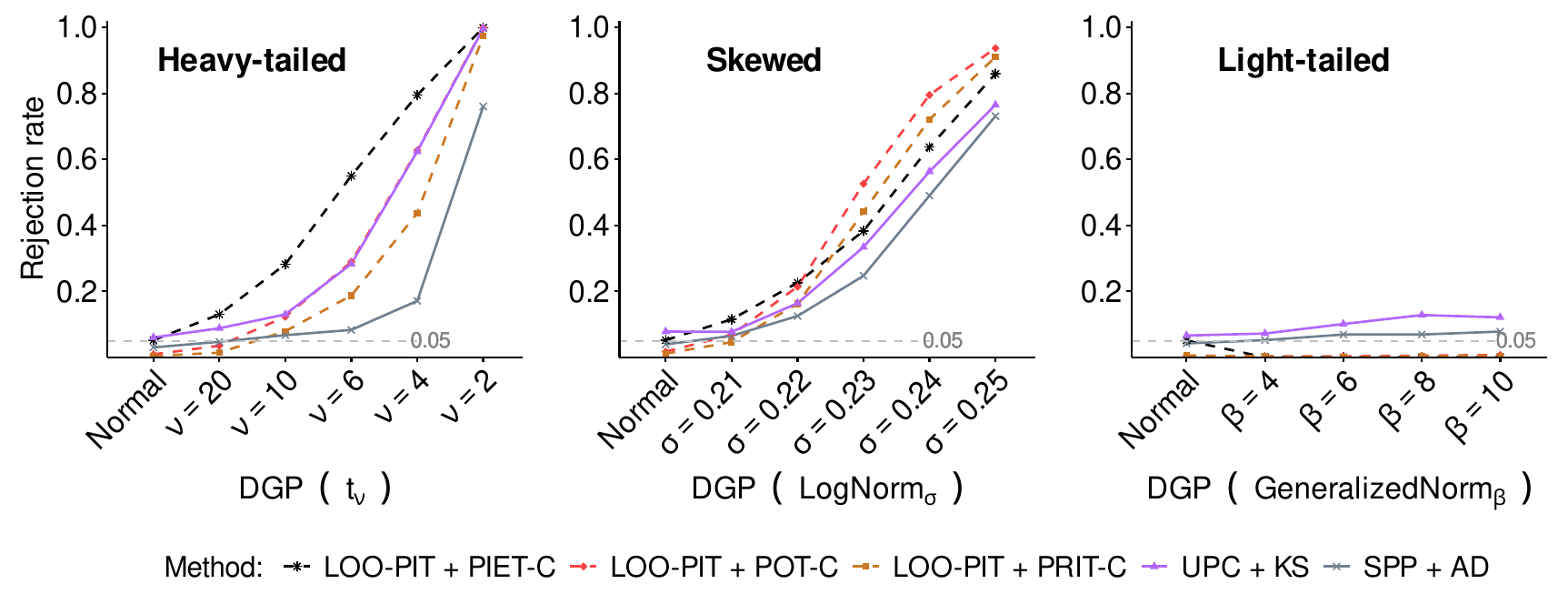} 
    \vspace{-16pt}
    \caption{Continuous model examples: Power performance evaluation of the tests under three scenarios for data distributions: Heavy-tailed, skewed and light-tailed. We always fit a normal model. x-axis shows true DGP. }
    \label{fig:2}
\end{figure}

 Results of the simulations are visualized in Figure~\ref{fig:2}. Our testing procedures generally outperform both the UPC and SPP methods. The improvement is most apparent for heavy-tailed data distributions, where the PIET-C provides a huge benefit in power. In this example, it seems the UPC and POT-C methods have similar performance, differing mainly in their Type I error rates, which are inflated and deflated, respectively. We also observe an advantage in using the proposed procedures when the data distributions are skewed. Interestingly, all tests have limited ability to detect departures involving lighter-than-normal tails. This limitation is particularly evident for the proposed tests, whose power drops to zero as the tail parameter of the generalized normal distribution increases beyond \(\beta > 2 \).
 This suggests the proposed methods may have difficulty detecting issues with a model when the true data distribution has thinner tails than the model-based distribution of the data. The PITOS method \citep{covington2025powerful} discussed in Appendix~\ref{sec:pitos} may be more effective at detecting this type of misspecification. See Appendix~\ref{sec:appdix_B} (Figure~\ref{fig:7}) for additional results with AD and KS tests. 
    
\paragraph{Hierarchical models for count data}
We extend the previous continuous-data experiments to discrete data models by using the randomized PIT values, as discussed by \citet{Czado-Gneiting-Held:2009}. Beyond the typical regime in which epistemic uncertainty mostly drives the model behavior, we also investigate a less common case where aleatoric variability (data noise) dominates in the posterior predictive distribution. Our findings indicate that SPP, when paired with the POT-C test, already offers a suitably calibrated check with competitive power in such cases. 

We start with a simple \textit{binomial} random effects model with weakly informative priors. the model structure is similar to before, 
\begin{align*}
y_{ig}  &\sim \operatorname{Binomial}\big(\ N,\, \mu_g \big)\\
\text{logit}( \mu_g) &\sim \mathcal{N}(\mu_0, \tau^2) \\
\mu_0 &\sim \mathcal{N}(0, 4^2), \ \ \tau \sim \mathcal{N}^+(0, 1)
\end{align*}
for all $i,g$. Meanwhile, under the true DGP, the data are generated according to a \textit{beta-binomial} model of the form, $y_{ig}\mid \mu_0,\tau,\phi\sim \operatorname{BetaBin}\big(N, \mu_g\cdot\phi, (1-\mu_g)\cdot\phi\big)$. We vary the concentration (overdispersion) parameter $\phi$ to control for the misspecification with respect to the binomial model. In fact, as $\phi$ increases, the latent success probability becomes tightly concentrated around its mean, and the \textit{beta–binomial} converges to \textit{binomial}. We used $\mu_0=\text{logit}^{-1}(0.4)$ and $\tau=1.5$. 

Similar experiment is conducted using an unbounded discrete distribution model---namely, a mixed-effects \textit{Poisson} model with the $\log$ link function: $y_{ig}  \sim \mathrm{Poisson}\big(\mu_g\big),\\ \log( \mu_g) \sim \mathcal{N}(\mu_0, \tau^2)\ $
for all $i,g$. The observed counts are generated from \textit{negative binomial}: $ y_{ig}\mid \mu_0,\tau,\phi \sim \mathrm{NegBin}\big(\mu_g,\phi)$, where departures from the Poisson assumption are introduced by allowing excess variability beyond the mean, controlled by the overdispersion parameter $\phi$. $\mu_0$ and $\tau$ were fixed to $\log(8)$ and $1.5$ respectively. 

 \begin{figure}[tp]
    \centering
    \includegraphics[width=\linewidth]{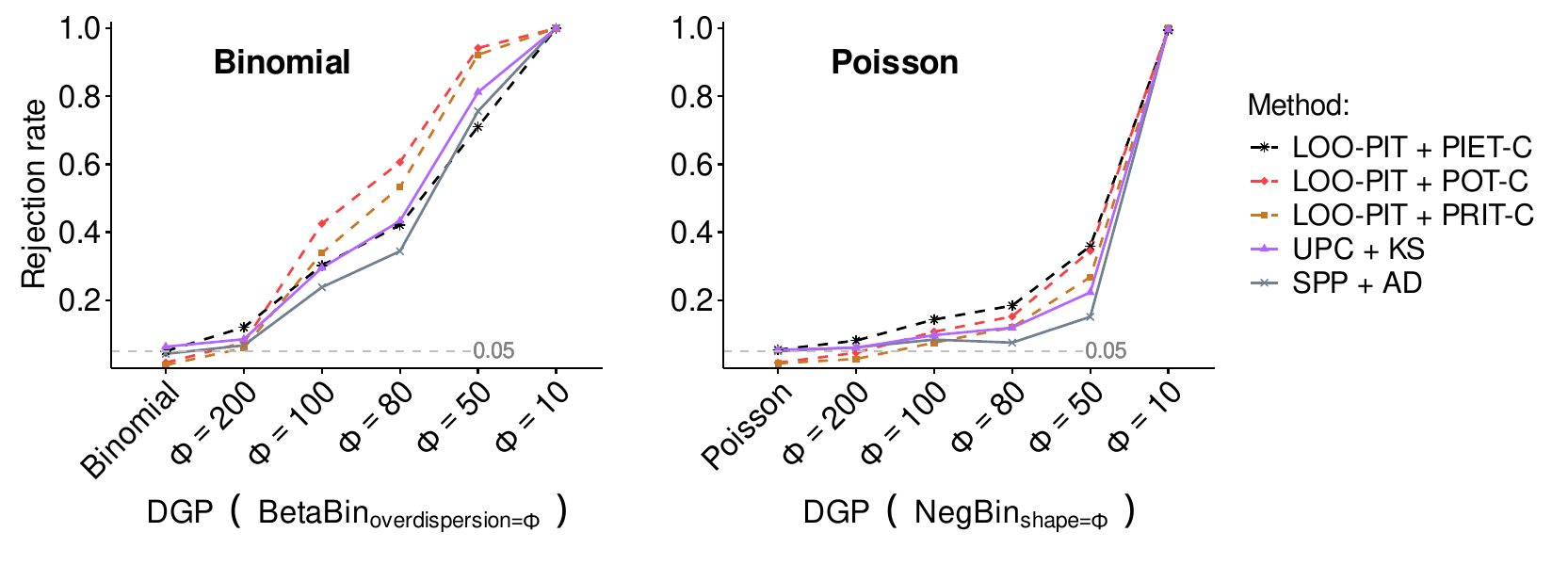}
    \vspace{-18pt}
    \caption{Discrete model examples: Power performance evaluation on binomial and Poisson models. x-axis are true DGP. Our methods generally improve upon UPC, although PIET-C shows reduced power in the binomial case.}
    \label{fig:3}
\end{figure}
Figure~\ref{fig:3} summarizes the simulation results. Overall, the proposed procedures again show better performance compared to considered competing methods. An exception arises in the binomial example, where PIET-C exhibits performance comparable to UPC, likely reflecting a modest loss of power due to the randomization required for discrete PIT values, since PIET-C is inherently more effective with continuous PITs. This effect is less evident in the Poisson example due to its near-continuous behavior.

\begin{figure}[tp]
    \centering    \includegraphics[width=\linewidth]{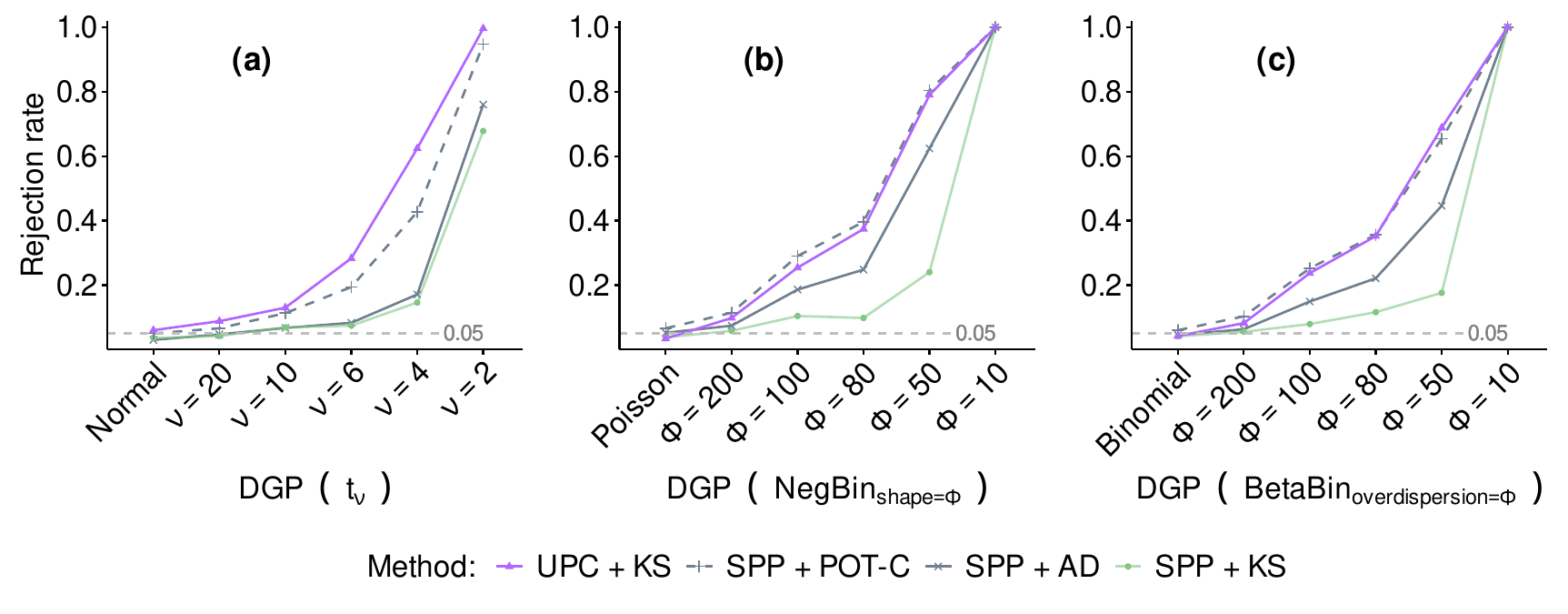}
    \vspace{-12pt}
    \caption{Power gain from using POT-C with SPP: \textbf{(a)} A typical scenario where model uncertainty matters significantly, POT-C enhances SPP power compared to AD and KS tests. \textbf{(b), (c)} Aleatoric uncertainty dominates, SPP becomes naturally more powerful and, when combined with POT-C, achieves performance comparable to UPC.}
    \label{fig:5}
\end{figure}

Both considered models and DGP are usual situations where the model uncertainty prevails in the posterior predictive variability. Next, we consider the scenario where intrinsic noise of the DGP dominates the epistemic uncertainty. In this context, if the posterior distribution is narrow, integrating over it has little effect on the predictive distribution, and using more than single posterior draw does not make much difference. For the same reason, it is likely that the LOO-posterior would not move much from the full posterior. An intuitive way to verify this is to compare the entropy of the predictive distribution obtained by averaging over the posterior with that computed using the posterior mean, $\Delta H=H\!\left(p(y|y^{\text{obs}})\right)-H\!\left(p(y \mid \hat{\theta})\right)$. If the difference is small, it probably suggests that aleatoric uncertainty is dominating, and thus the posterior is not sensitive to leave-one-out perturbations. We empirically show with a real data example that SPP approach might be a computationally preferable choice in such cases (see Figure~\ref{fig:5}). 

\begin{figure}[tp]
    \centering
    \includegraphics[width=\linewidth]{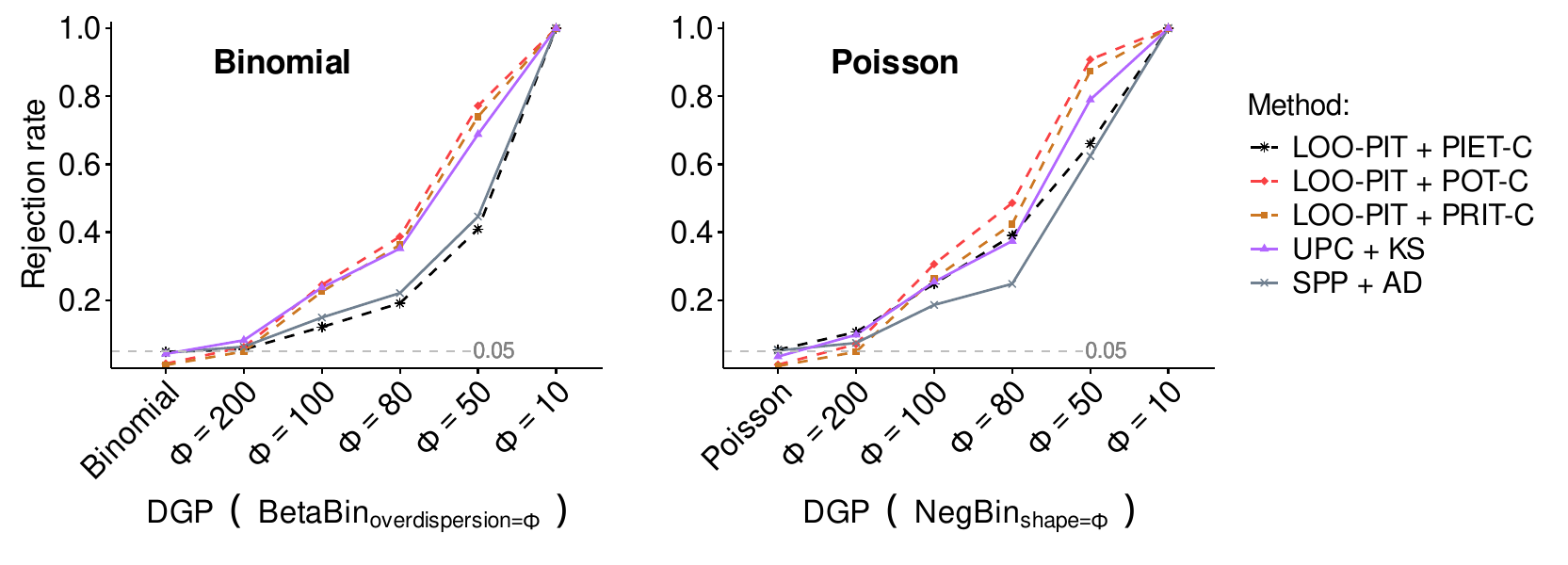}
    \vspace{-18pt}
    \caption{Power performance when aleatoric uncertainty is dominating: UPC and often SPP methods generally have some advantages over PIET-C in this situation, but again with slightly lower power compared to POT-C procedure.}
    \label{fig:4}
\end{figure}

 The dataset used comes from a clinical study analysed by \cite{lintzeris2019nabiximols}. 
It comprises data from 128 participants ($\mathsf{id}$) randomized into two treatment groups ($\mathsf{group}$ = Placebo or Nabiximols). Participants received $12$-week treatment involving weekly clinical reviews. The primary outcome was the self-reported number of illicit cannabis used days ($\mathsf{y}$ ) for previous $28$ day ($\mathsf{trials}$), asked after  $0, 4, 8,$ and $12$ weeks ($\mathsf{week}$ as factor). The data exhibit group-level imbalance, with unequal numbers of measurement across treatment groups. Upon removing missing follow-up visit, the total number of observations is $n=385$.  The data have been used to illustrate various aspects of Bayesian workflow, including  comparison of models \citep{Gelman-Vehtari-Mcelreath-etal:2026}. In particular a \textit{beta-binomial} with interaction terms ($\mathsf{group}*\mathsf{week}$) and varying intercept for each participant ($\mathsf{id}$) has been shown to provide a good fit to the data. Specifically, the model has the following form with specified priors: $\mathsf{y}_{ik}|\mathsf{\beta},\tau,\phi\sim \operatorname{BetaBin}(\mathsf{trials},\mu_{ik}.\phi, (1-\mu_{ik}).\phi)$, where $\mu_{ik}=\text{logit}^{-1}(\mu_i+ \beta_1 \mathsf{group}_{ik}+\beta_2 \mathsf{week}_{ik}+\beta_3\mathsf{group}_{ik}\times \mathsf{week}_{ik})$  for $i\in\{1,\cdots,128\}, k\in\{0,4,8,12\}$ . In this case, the estimated $\hat{p}_{\loo}$ is quite high. In addition, this example is characterized by the aleatoric part driving the model’s behavior, as further confirmed by the entropy analysis described above. 
 
For the purpose of our study, we copy this property of the Nabiximols's data by generating comparable responses ($\mathsf{y}$) using the posterior mean of the latter \textit{beta-binomial} model for Nabiximols' data, while varying the overdispersion. Meanwhile, just as before, we assume a binomial model with participant-specific varying intercept, but now conditioning on additional covariates $\mathsf{week}$ and $\mathsf{group}$.  Slightly wider priors are used, especially as the data seem to be quite informative.
\begin{align*}
\mathsf{y}_{ik}  &\sim \mathrm{Binomial}\big(\mathsf{trials},\, \text{logit}^{-1}(\mu_i+ \beta_1 \mathsf{group}_{ik}+\beta_2 \mathsf{week}_{ik}+\beta_3\mathsf{group}_{ik}\times \mathsf{week}_{ik})\big),\\
\mu_i&\sim \mathcal{N}(\mu_0, \tau^2),\\
\mu_0 &\sim \mathcal{N}(0, 5^2),\ \ \bm{\beta}\sim \mathcal{N}(0, 5^2),\ \ \tau \sim \mathcal{N}^+(0, 1.5^2).
\end{align*}
 Similarly, we further examine the case where true DGP follows a negative binomial model calibrated to the Nabiximols' data, while inference is conducted under a Poisson assumption with same priors: 
\begin{align*}
\mathsf{y}_{ik} \sim \mathrm{Poisson}\left(\exp(\mu_i+ \beta_1 \mathsf{group}_{ik}+\beta_2 \mathsf{week}_{ik}+\beta_3\mathsf{group}_{ik}\times \mathsf{week}_{ik})\right),\ 
\mu_i\sim \mathcal{N}(\mu_0, \tau^2).
\end{align*}
 
Figure~\ref{fig:4} shows that under this situation, the UPC approach and often SPP (in the Binomial case ) demonstrate clearly better performance than PIET-C, although they have slightly less power than POT-C (and PRIT-C) but provide better Type I error control. In addition, Figure~\ref{fig:5} provides visual evidence that combining SPP with POT-C as the uniformity check yields improved power. In particular, Figure~\ref{fig:5}~\textbf{(a)} shows the gain in power achieved when assessing the normal model using SPP with POT-C (SPP+POT-C), compared to AD and KS tests. Also, in the present setting, where aleatoric uncertainty dominates, it turns out that SPP+UPC performs comparably to the UPC method, as visualized in Figure~\ref{fig:5}~\textbf{(b)} and \textbf{(c)}. This suggests that SPP+POT-C may be preferable over UPC and LOO-PIT diagnostics in such cases, at least for its efficiency. More generally, this result highlights the usefulness of POT-C as a powerful uniformity test, even under independence assumptions (see Figure~\ref{fig:power_an} in Appendix~\ref{sec:appdix_B} for power analysis study).
\section{Conclusion}
\label{sec:discussion}
We have introduced POT-C, PRIT-C, and PIET-C testing procedures that bring together distributional theory for pointwise uniformity checks with Cauchy methods for dependent $p$-values aggregation. The motivation for developing such tests, in the context of Bayesian model criticism, was to address the fold-induced dependence present in LOO-PIT values, which would lower the power of classical uniformity tests based on the independence assumption. We established theoretical results on the origin of the LOO-PIT dependence, and demonstrated that its strength depends on how complex the model is relative to the sample size. With the exception of PIET-C, which maintains valid Type I error control, the calibration of POT-C and PRIT-C depends on the strength of the underlying dependence structure. In terms of power to detect model miscalibration, we demonstrated through examples that the proposed tests are competitive and often substantially more powerful than competing methods considered. Finally, we introduce a graphical procedure that is more sensitive and computationally more efficient than the commonly used ECDF-based graphical uniformity test based on the independence assumption \citep{sailynoja2022graphical}. Although we focused on testing LOO-PIT values, the developed methods. 

We recommend computing PIT values using distribution CDFs when available, and an iterative process to PIT-based model checking. We can start with SPP, the computationally simplest method. If the model passes SPP but LOO computation is too expensive or fails, UPC can be used next. Otherwise, consider using LOO-PIT diagnostics per Table~\ref{tab:sum}. At any level, if the PIT values flag a misfit, further tests are unnecessary. For testing LOO-PIT values, we suggest POT-C test as the default choice due to its good power against diverse type of local departures from the null. The PIET-C test focuses specifically on tail deviations and PRIT-C is weaker than POT-C for LOO-PITs.

Future research may explore how to explicitly describe the dependency structure in LOO-PIT values and adjust simultaneous confidence bands for the graphical ECDF test accordingly---make them narrower to achieve calibration comparable or eventually superior to that of our actual tests.

\begin{table}[tp]
\small
\centering
\renewcommand{\arraystretch}{1.4}
\begin{tabular}{p{1.3cm} p{2.8cm} p{3cm} p{5cm}}

\textbf{Method} & \textbf{Pointwise basis} & \textbf{PITs nature} & \textbf{Recommended use} \\
\midrule
POT-C  
& beta (marginal \newline  order statistics) 
& continuous \newline PIT values 
& Should ideally use continuous LOO-PITs as computed using the parametric observation CDF given posterior draws. Works for PSIS-LOO weighted rank-based PITs because the continuously-varying importance weights produce LOO-PIT values that behave as approximately continuous in practice, albeit potential inaccuracy in the extreme tails. \\
\midrule
PRIT-C 
& binomial \newline (scaled ECDF) 
& discrete or rank\newline -based PIT values 
& Mostly compatible with PITs computed as normalized ranks. Weaker than POT-C in testing weighted rank based LOO-PIT values for the same reason previously outlined (LOO-PIT values look continuous).\\
\midrule
PIET-C 
& continuous inverse-CDF transform
& continuous \newline PIT  values 
& Exclusively recommended for detecting deviations in tails (or outliers). Continuous PIT values should be used to guarantee valid Type I error control under dependence. \\
\hline
\end{tabular}
\caption{Summary of proposed testing procedures with guidance on their appropriate use for LOO-PIT-based model checking. When using POT-C and PIET-C approaches, we recommend computing PIT values using the known parametric observation CDF whenever possible, as this improves accuracy in tails (e.g., avoids PIT values of exactly $0$ and $1$) and reduces variability in PIT uniformity checking.}
\label{tab:sum}
\end{table}

\section*{Acknowledgements}

We thank David Kohns and Osvaldo Martin for helpful comments, and Florence Bockting for  assistance with the implementation in the bayesplot R package. The research has been supported by the Research Council of Finland Flagship programme: Finnish Center for Artificial Intelligence, and Research Council of Finland grants (340721, 368069). This work was also a part of Finland's Ministry of Education and Culture’s Doctoral Education Pilot under Decision No. VN/3137/2024-OKM-6 (The Finnish Doctoral Program Network in Artificial Intelligence, AI-DOC).

{
\small
\bibliography{references}
}


\appendix
\section{ Supporting Derivations}
\subsection{Posterior PIT in terms of predictive CDF}
\label{sec:appdix_A}
\begin{proof} 

    \begin{align*}
  p_{\mathrm{post},i}&=\int_{-\infty}^{y_i}P(\tilde{y}_i|\bm{y}) d\tilde{y}_i\\
  &=\int_{-\infty}^{y_i} \int_{\Theta} p( \tilde{y}_i|\x_i,\theta)\pi(\bm{\theta}|\bm{y})d\bm{\theta}d\tilde{y}_i\\
  &=\int_{\Theta} \Big(\int_{-\infty}^{y_i}  p( \tilde{y}_i|\x_i,\theta)d\tilde{y}_i \Big) \pi(\bm{\theta}|\bm{y})d\bm{\theta}\\
  &=\int F_i(\bm{\theta}) \pi(\bm{\theta}|\bm{y})d\bm{\theta}
\end{align*}
\end{proof}

\subsection{ Taylor expansion and integration against the posterior}
\label{sec:appdix_A2}
\begin{proof} 

$$F_i(\theta)=F_i(\bm{\theta}_0)
+ \nabla_{\bm{\theta}} F_i(\bm{\theta}_0)^{\top}(\bm{\theta} - \bm{\theta}_0)+ 
\frac{1}{2}(\bm{\theta} - \bm{\theta}_0)^{\top}
\nabla^2_\theta F_i(\bm{\theta}_0)(\bm{\theta} - \bm{\theta}_0)+O(\parallel \theta-\theta_0\parallel^3)
$$

Integrating both sides with respect to the full data posterior $\pi_n:=\pi(\bm{\theta}|\mathcal{D}_n)$ and using BvM theorem for the asymptotic posterior's covariance and expectation,
\begin{align*}
    p_{\mathrm{post},i}&=\int F_i(\bm{\theta}) \pi(\bm{\theta}|\mathcal{D}_n)d\bm{\theta}\\
    &=\mathbb{E}_{\pi_n}\left[F_i(\bm{\theta})\right]\\
    &= F_i(\bm{\theta}_0)
+ \nabla_{\bm{\theta}} F_i(\bm{\theta}_0)^{\top}(\overbrace{\mathbb{E}_{\pi_n}\left[\bm{\theta}\right]}^{\hat{\theta}_n} - \bm{\theta}_0)\\
&+ 
\frac{1}{2}\mathbb{E}_{\pi_n}\left[(\bm{\theta} - \bm{\theta}_0)^{\top}
\nabla^2_\theta F_i(\bm{\theta}_0)(\bm{\theta} - \bm{\theta}_0)\right] +O\left(\mathbb{E}_{\pi_n}\left[\parallel \theta-\theta_0\parallel^3\right]\right).
\end{align*}


\paragraph{Second-order expectation.}
\begin{align*}
    \mathbb{E}_{\pi_n}\big[(\bm{\theta} - \bm{\theta}_0)^{\top} \nabla^2_\theta F_i(\bm{\theta}_0)(\bm{\theta} - \bm{\theta}_0)\big]&=
    \mathbb{E}_{\pi_n}\big[tr\big\{(\bm{\theta} - \bm{\theta}_0)^{\top} \nabla^2_\theta F_i(\bm{\theta}_0)(\bm{\theta} - \bm{\theta}_0)\big\}\big]\\
    &=\mathbb{E}_{\pi_n}\big[tr\big\{\nabla^2_\theta F_i(\tilde{\bm{\theta}})(\bm{\theta} - \bm{\theta}_0)(\bm{\theta} - \bm{\theta}_0)^{\top} \big\}\big]\\
    &=tr\big\{\nabla^2_\theta F_i(\bm{\theta}_0)\mathbb{E}_{\pi_n}\big[(\bm{\theta} - \bm{\theta}_0)(\bm{\theta} - \bm{\theta}_0)^{\top} \big]\big\}\\
    &\approx tr\big\{\nabla^2_\theta F_i(\bm{\theta}_0).\frac{1}{n}\mathcal{I}^{-1}(\bm{\theta}_0)\big\}\\
    &=\frac{1}{n}tr\big\{\nabla^2_\theta F_i(\bm{\theta}_0)\mathcal{I}^{-1}(\bm{\theta}_0)\big\}
\end{align*}

\paragraph{Remainder term. }

We want to show that
\[
\mathbb{E}\|\theta-\theta_0\|^3=O(n^{-3/2}).
\]
Asymptotically, under BvM conditions, the have that

\[
\theta-\theta_0=\hat{\theta}_n-\theta_0+\frac{1}{\sqrt{n}}LZ,
\qquad
Z\sim\mathcal{N}(0,I_p),
\qquad
LL^{\top}=I^{-1}(\theta_0),
\]
where $L$ is the Cholesky factor of $I^{-1}(\theta_0)$. Write
$
X=\hat{\theta}_n-\theta_0,\ \text{and}\
Y=\frac{1}{\sqrt{n}}LZ, $
so that \(\theta-\theta_0=X+Y\). By Minkowski's inequality in \(L^3\),

$$\ \left(\mathbb{E}\|X+Y\|^3\right)^{1/3}
\leq
\left(\mathbb{E}\|X\|^3\right)^{1/3}
+
\left(\mathbb{E}\|Y\|^3\right)^{1/3}.
$$

Therefore,

\[
\left(\mathbb{E}\|\theta-\theta_0\|^3\right)^{1/3}
\leq
\left(\mathbb{E}\|\hat{\theta}_n-\theta_0\|^3\right)^{1/3}
+
\left(
\mathbb{E}
\left\|
\frac{1}{\sqrt{n}}LZ
\right\|^3
\right)^{1/3}.
\]

Using the homogeneity of the norm and linearity of the expectation,
\[
\mathbb{E}
\left\|
\frac{1}{\sqrt{n}}LZ
\right\|^3
=
n^{-3/2}\mathbb{E}\|LZ\|^3.
\]

hence
\begin{align*}
    \left(
\mathbb{E}
\left\|
\frac{1}{\sqrt{n}}LZ
\right\|^3
\right)^{1/3}
&=
n^{-1/2}
\left(
\mathbb{E}\|LZ\|^3
\right)^{1/3}\\
&=O(n^{-1/2}),\ \ \ \text{since}\ \
\mathbb{E}\|LZ\|^3<\infty.
\end{align*}

Similarly, 
$$ \left(\mathbb{E}_{\pi_n}\|\hat{\theta}_n-\theta_0\|^3\right)^{1/3}=O(n^{-1/2}).$$
This follows from the asymptotic property of the MLE, $\sqrt{n}(\hat{\bm{\theta}}_n-\bm{\theta}_0) \xrightarrow{d} \mathcal{N}\big(0,\mathcal{I}^{-1}(\bm{\theta}_0 )\big),$ and a same Cholesky based decomposition of $\hat{\bm{\theta}}_n-\bm{\theta}_0$ as before.

We then obtain,
\[
\left(\mathbb{E}\|\theta-\theta_0\|^3\right)^{1/3}\leq O(n^{-1/2})+O(n^{-1/2})=O(n^{-1/2}).
\]
Therefore,
\[\left(\mathbb{E}\|\theta-\theta_0\|^3\right)^{1/3}=O(n^{-1/2})\]

Cubing both sides yields
\[
\boxed{
\mathbb{E}\|\theta-\theta_0\|^3=O(n^{-3/2})
}.
\]
\end{proof}

\subsection{ The self-influence terms are uncorrelated under i.i.d. sampling.}
\label{sec:appdix_A3}
\begin{proof}

\begin{equation*}
    c_i = \nabla_{\bm{\theta}} F_i(\bm{\theta}_0)^\top \IF(y_i), \qquad 
    \IF(y_i) = \mathcal{H}_n^{-1}(\hat{\bm{\theta}}_n)\, \nabla_{\bm{\theta}} \log p(y_i \mid \mathbf{x}_i, \hat{\bm{\theta}}_n).
\end{equation*}
Conditional on $\hat{\bm{\theta}}_n$, $c_i$ is a function of $y_i$ alone. Given the $\{y_i\}$ are i.i.d., it follows that
    $c_i \perp c_j \mid \hat{\bm{\theta}}_n,\ i \neq j, $
and in particular, $\mathrm{Cov}(c_i, c_j \mid \hat{\bm{\theta}}_n) = 0$ for all $i \neq j$.

By the law of total covariance, for $i \neq j$:
\begin{equation*}
    \mathrm{Cov}(c_i, c_j) = 
    \underbrace{\mathbb{E}\left[\mathrm{Cov}(c_i, c_j \mid \hat{\bm{\theta}}_n)\right]}_{=\, 0 } 
    + \mathrm{Cov}\left(\mathbb{E}[c_i \mid \hat{\bm{\theta}}_n],\, \mathbb{E}[c_j \mid \hat{\bm{\theta}}_n]\right).
\end{equation*}
It remains to show that the second term vanishes. The conditional expectation of $c_i$ given $\hat{\bm{\theta}}_n$ is
\[   
\mathbb{E}[c_i \mid \hat{\bm{\theta}}_n] = \nabla_{\bm{\theta}} F_i(\ \bm{\theta}_0)^\top \mathcal{H}_n^{-1}(\hat{\bm{\theta}}_n)\, 
    \mathbb{E}\left[\nabla_{\bm{\theta}} \log p(y_i \mid \mathbf{x}_i, \hat{\bm{\theta}}_n) \mid \hat{\bm{\theta}}_n\right].
\]
Now, because $\hat{\bm{\theta}}_n$ is the MLE of $\theta$, it satisfies the score equation
$$\sum_{i=1}^n \nabla_{\bm{\theta}} \log p(y_i \mid \mathbf{x}_i, \hat{\bm{\theta}}_n) = 0\quad \text{ almost surely}.$$
Taking conditional expectations on both sides given  $\hat{\bm{\theta}}_n$ and using the linearity of expectation,
$$
    \sum_{i=1}^n \mathbb{E}\left[\nabla_{\bm{\theta}} \log p(y_i \mid \mathbf{x}_i, \hat{\bm{\theta}}_n)\right] = 0.
$$
Under i.i.d.\ setting, the conditional distribution of each $y_i$
given $\hat{\bm{\theta}}_n$ is the same, and hence each term in the sum has the same conditional expectation. The equation above then becomes
$   n\, \mathbb{E}\left[\nabla_{\bm{\theta}} \log p(y_i \mid \mathbf{x}_i, \hat{\bm{\theta}}_n) \mid \hat{\bm{\theta}}_n\right] = 0\Longrightarrow \mathbb{E}\left[\nabla_{\bm{\theta}} \log p(y_i \mid \mathbf{x}_i, \hat{\bm{\theta}}_n) \mid \hat{\bm{\theta}}_n\right] = 0.$

Hence $\mathbb{E}[c_i \mid \hat{\bm{\theta}}_n] = 0$ for all $i$, and thus $ \mathrm{Cov}\left(\mathbb{E}[c_i \mid \hat{\bm{\theta}}_n],\, \mathbb{E}[c_j \mid \hat{\bm{\theta}}_n]\right) =  0.$
Therefore, $$\mathrm{Cov}(c_i, c_j) = 0\ \ \text{for all}\ \ i \neq j.$$
\end{proof}

\subsection{ Expected marginal contributions to the overall test }
\label{sec:appdix_A4}
\begin{proof} 
    For any coalition $S \subseteq N \setminus \{u_i\}$,  let us denote by $s=|S|$ its cardinality. Hereafter, we identify each ``\textit{player}'' $u_{i}$ by its index $i$, for $i=1,...,n$.
We first reorganize the Shapley values formulation by grouping coalitions by size. That is, instead of summing over all subsets in one big sum, we do:  \\

1. Outer sum: loops over subset (coalition set) sizes, $s=0,1,...,n-1$

2. Inner sum: sum marginal contributions over all subsets of exactly that size.

Since the weight depends only on $s=|S|$ it can be pulled outside the inner sum.\\
  \begin{align*}
\phi_i(v) 
&= \sum_{S \subseteq N \setminus \{i\}}
\frac{|S|!\,(n - |S| - 1)!}{n!}
\big( v(S \cup \{i\}) - v(S) \big) \\
&= \sum_{s=0}^{n-1}
\frac{s!\,(n - s - 1)!}{n!}
\sum_{ |S| = s}
\big\{ v(S \cup \{i\}) - v(S) \big\}
\end{align*}

\begin{itemize}
    \item Marginal contribution: the contribution of the "player" $u_i$ to the coalition $S$ is given by,
    $$ v(S \cup \{i\}) - v(S)
= \frac{1}{s+1}\big(\sum_{j\in S} t_j + t_i\big) - \frac{1}{s}\sum_{ j\in S} t_j
= \frac{t_i - \overline{t}_S}{s+1}$$
where $\overline{t}_S = \frac{1}{s}\sum_{S} t_j$.
For $s=0$ (i.e., coalition of no players), the marginal contribution is simply $v(\{u_{(i)}\}) - v(\varnothing) = t_i$.

\item Aggregation by subset size: for fixed $s \ge 1$, the number of subsets $S \subseteq N \setminus \{i\}$ of size $s$ is $\binom{n-1}{s}$.
The total contribution over all coalitions of size $s$ is then, 
\begin{align*}
    \sum_{ |S| = s}
\big\{ v(S \cup \{i\}) - v(S) \big\}
&= \sum_{|S| = s} \frac{t_i - \overline{t}_S}{s+1}\\
&=\frac{1}{s+1}\Big[ \sum_{|S| = s} t_i - \sum_{|S| = s} \overline{t}_S \Big]\\
&= \frac{1}{s+1}\Big[ \binom{n-1}{s}t_i - \sum_{|S| = s} \overline{t}_S \Big].
\end{align*}

Now, calculate 
\[
\sum_{|S| = s}\overline{t}_S
= \sum_{|S| = s}\frac{1}{s}\sum_{S} t_j
= \frac{1}{s}\sum_{j\neq i }t_j \cdot \#\{S : |S|=s,\, j \in S\}.
\]

What does it mean ? : Instead of summing over subsets first, we sum over elements first. Each player $j$ appears in many coalitions $S$ of size $s$. 
For fixed $j\neq i$, the number of such coalitions is, $\#\{S : |S|=s,\, j \in S\}=\binom{n-2}{s-1}$. Hence
\[
\sum_{|S| = s}\overline{t}_S = \frac{1}{s}\,\binom{n-2}{s-1}\,t_{_{-i}}\ ,
\]
where $t_{_{-i}}=\sum_{j \neq i }t_j $ is the sum of all players’ values except $t_i$. Therefore,
$$
 \sum_{ |S| = s}
\big\{ v(S \cup \{u_{(i)}\}) - v(S) \big\}
= \frac{1}{s+1}\!\left[\binom{n-1}{s}t_i - \frac{1}{s}\binom{n-2}{s-1}t_{_{-i}}\right].
$$

\item Apply the combinatorial weight: the contribution from all coalitions of size $s\geq 1$ is
\begin{align*}
  & \frac{s!(n-s-1)!}{n!} \sum_{ |S| = s}
\big\{ v(S \cup \{i\}) - v(S) \big\}
= \frac{s!(n-s-1)!}{n!}\cdot \frac{1}{s+1}\!\left[\binom{n-1}{s}t_i - \frac{1}{s}\binom{n-2}{s-1}t_{_{-i}}\right]\\ 
&=\frac{s!(n-s-1)!}{n!}\cdot \frac{1}{s+1}\!\left[\frac{s!(n-s-1)!}{n!}\cdot\frac{(n-1)!}{s!(n-1-s)!}t_i - \frac{1}{n!}\cdot\frac{s!(n-s-1)!}{s}\cdot\frac{(n-2)!}{(s-1)!(n-s-1)!}t_{_{-i}}\right]\\
&= \frac{1}{s+1} \left[ \frac{1}{n}\cdot t_i  - \frac{1}{n(n-1)}\cdot t_{_{-i}}\right] 
\end{align*}

\item{ Summing over coalition sizes}: for $s=0$, the weight is $\frac{(n-1)!}{n!}=\frac{1}{n}$ and the marginal contribution is $t_i$, so the contribution is $\frac{1}{n}t_i$.

\begin{align*}
    \phi_i(v)&=\sum_{s=0}^{n-1}
\frac{s!\,(n - s - 1)!}{n!}
\sum_{ |S| = s}
\big\{ v(S \cup \{i\}) - v(S) \big\}\\
&= \frac{1}{n}t_i + \sum_{s=1}^{n-1}\frac{1}{s+1} \left[ \frac{1}{n}\cdot t_i  - \frac{1}{n(n-1)}\cdot t_{_{-i}}\right] \\
&=\frac{1}{n}t_i +\frac{1}{n} \bigg(t_i  - \frac{t_{_{-i}}}{(n-1)} \bigg)\sum_{s=1}^{n-1}\frac{1}{s+1} \\
&=\frac{1}{n}t_i +\frac{1}{n} \left[t_i  - \frac{t_{_{-i}}}{(n-1)} \right]\left(\sum_{k=1}^n \frac{1}{k}\ \ -\ 1\right)
\end{align*}

Thus ,
$$\boxed{
\phi_i(v) 
= \frac{1}{n}t_i + \frac{H_n - 1}{n}\bigg(t_i - \frac{1}{n-1}\sum_{j \neq i} t_j\bigg)
}\ ,$$
with 
 $\ \ t_i=tan\{(0.5-p^{(i)})\pi\}, \ and\quad H_n = \sum_{k=1}^n\frac{1}{k}\ \ $ ( the  Harmonic number ).

\end{itemize}
  
\end{proof}

\section{Additional results}

\label{sec:appdix_B}
\begin{figure}[tp]
    \centering    \includegraphics[width=\linewidth]{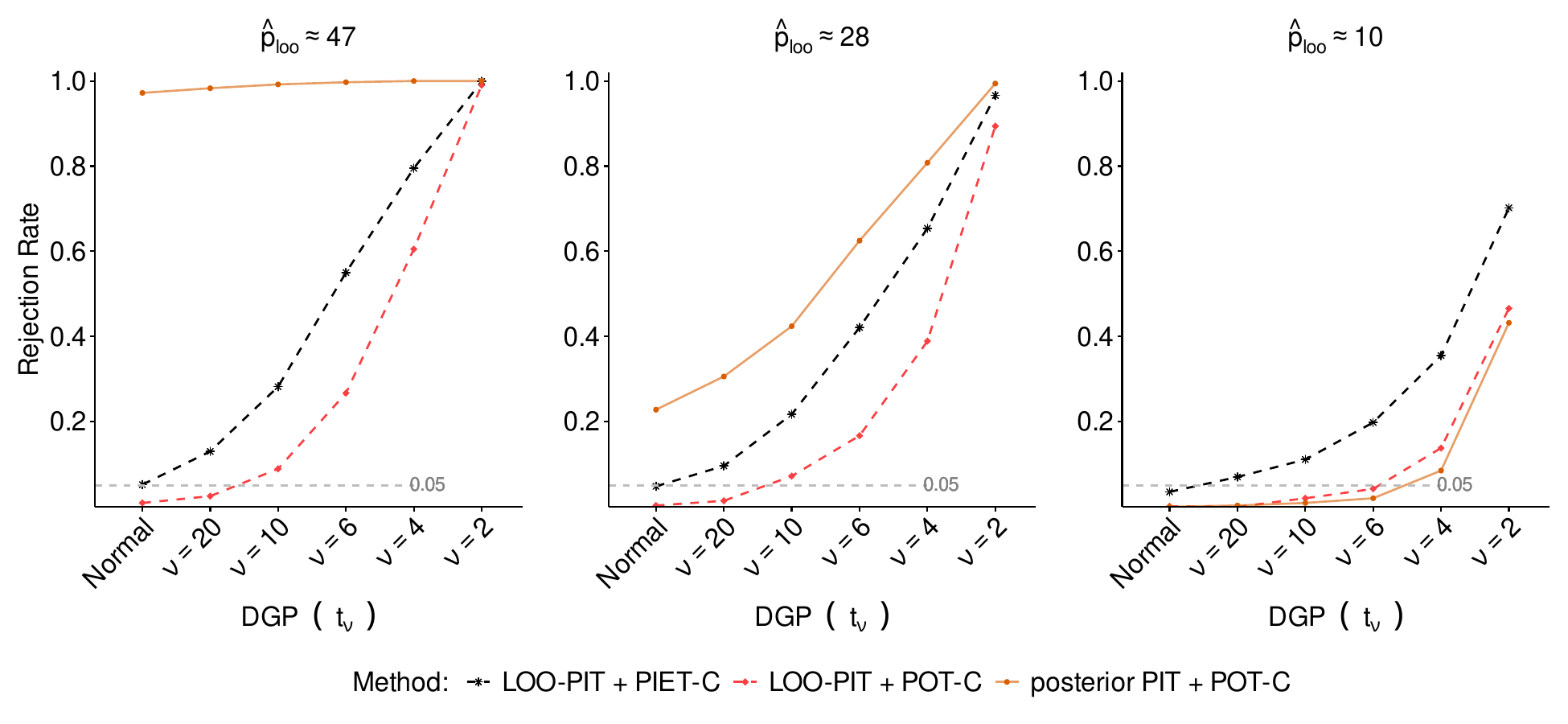}
    \caption{ Normal vs. Student $t$ example illustrating the increasing agreement between posterior PIT and LOO-PIT as model complexity reduces together with the sample size. Keeping \#observations per group to $m = 5$, we progressively reduce \#groups to $G=50$ (left), $G=30$ (middle), and $G=10$ (right), which naturally decreases $\hat{p}_{\mathrm{loo}}$. This is to clarify the fact that posterior PIT approaching LOO-PIT does not necessarily mean that dependence has vanished (except in the asymptotic regime), since the ratio $\hat{p}_{\mathrm{loo}}/n$ remains non-negligible in all cases. }
    \label{fig:8}
\end{figure}

\begin{figure}[tp]
    \centering    \includegraphics[width=1\linewidth]{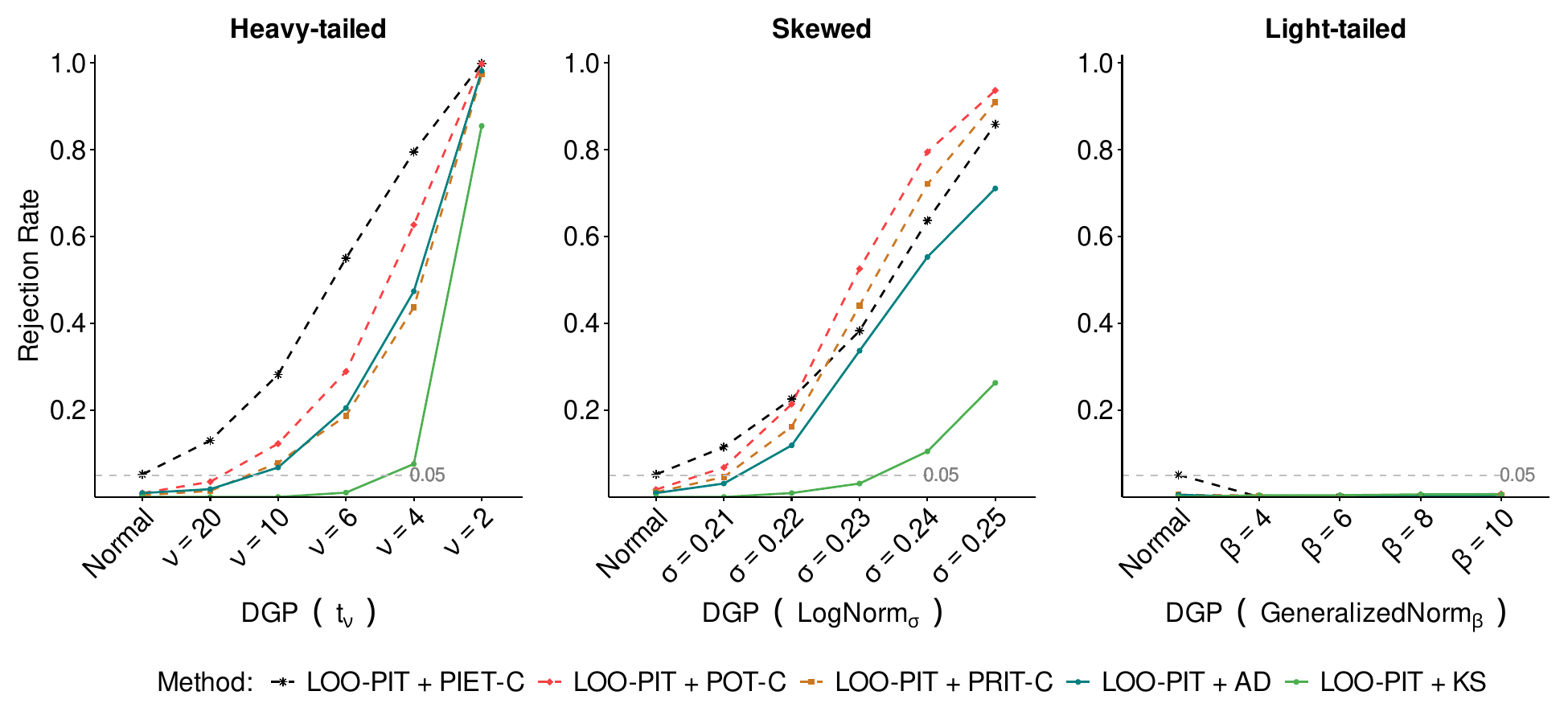}
    \caption{Continuous model examples: Power performance comparison with AD and KS tests under three scenarios for data distributions : Heavy-tailed, skewed and light-tailed. We always fit a normal model. x-axis are true DGP.
    POT-C and PIET-C are substantially more powerful than AD and KS, both of
whom originally assume the LOO-PIT values are independent. In particular, KS is poorly suited for detecting deviations in the tails. Again, lighter than normal tails make it difficult for the tests to detect issues with the normal model. }
    \label{fig:7}
\end{figure}
\begin{figure}[tp]
    \centering    \includegraphics[width=\linewidth]{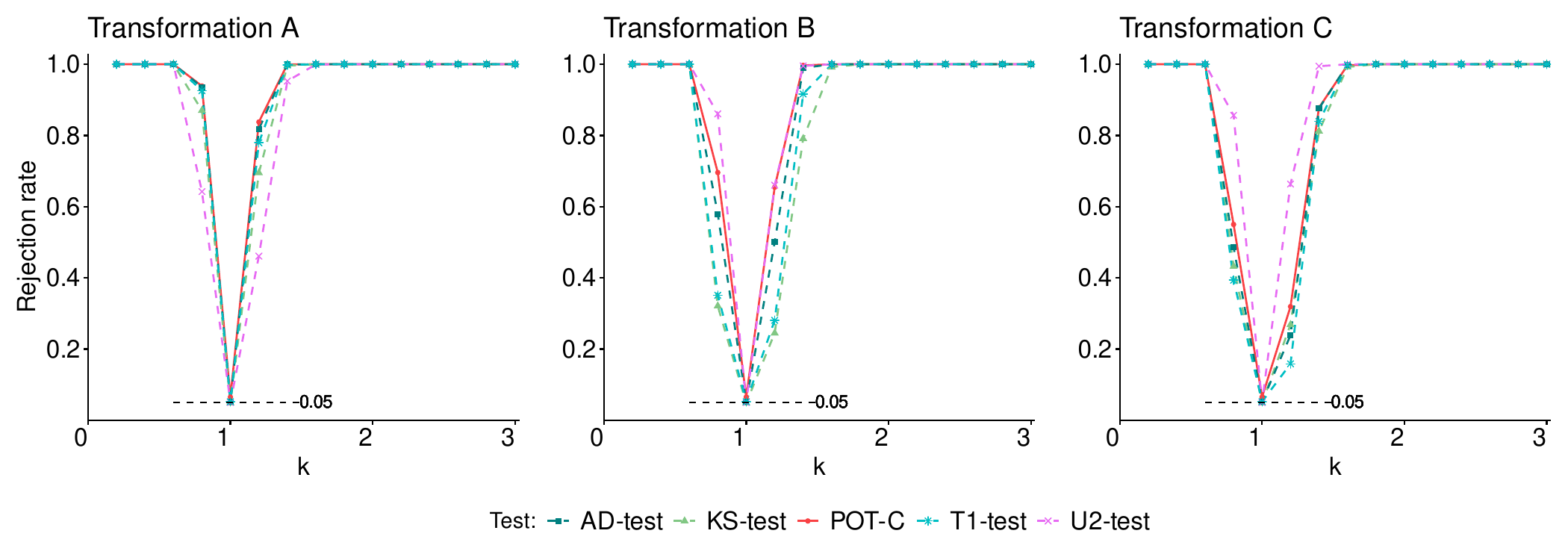}
    \caption{ Power analysis under independence. When compared to existing tests for uniformity, the POT-C test (red curve) shows strong rejection performance across all three families of deviations \citet{marhuenda2005comparison}  use  comparing various tests for uniformity. Rejection rates are based on $100{,}000$ samples for each value of $k$.}
 \label{fig:power_an}
\end{figure}

\newpage

\section{ Probability integral
transform of order statistics (PITOS) }
\label{sec:pitos}
\begin{figure}[tp]
    \centering    \includegraphics[width=\linewidth]{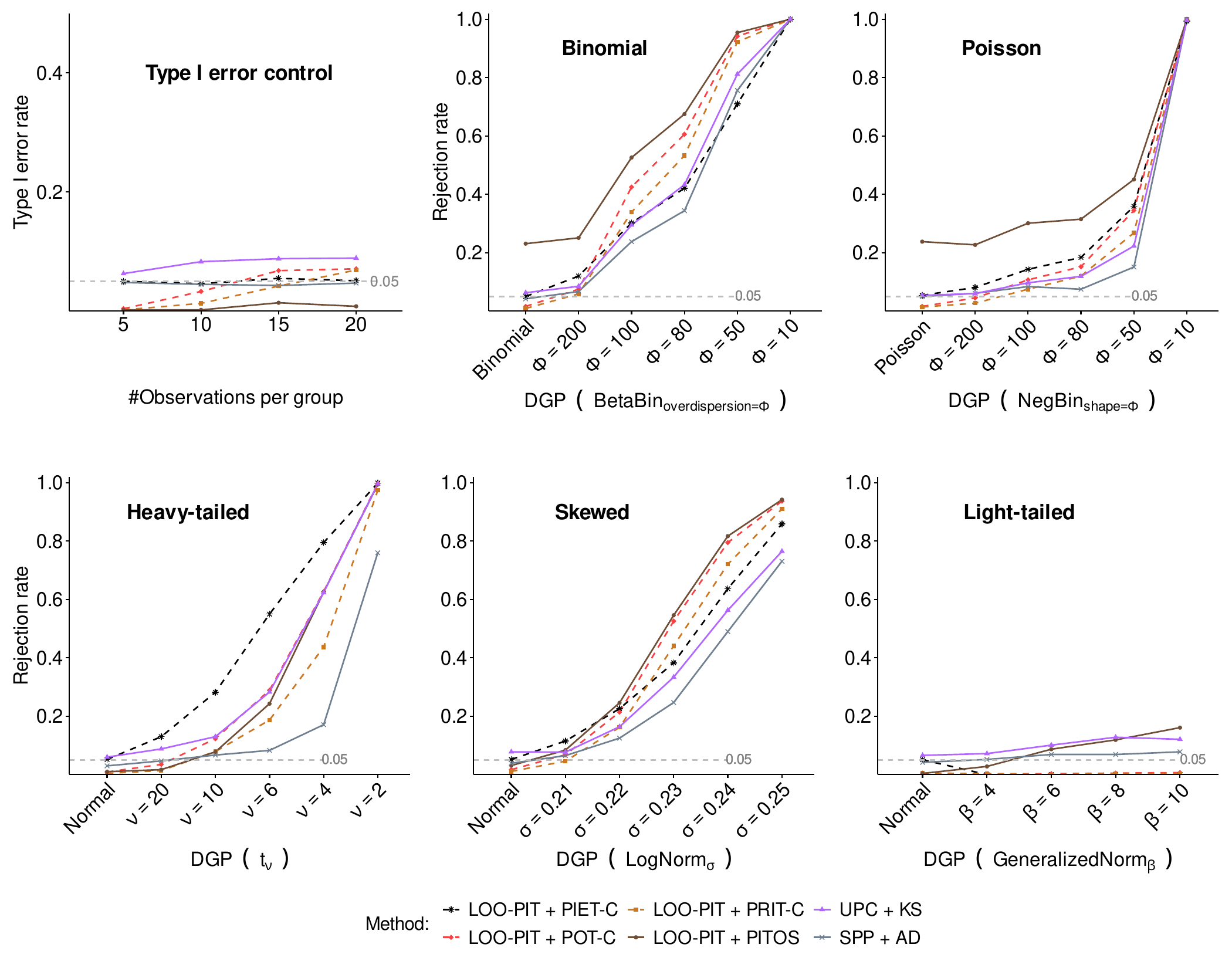}
    \caption{PITOS performance across the studied examples. While showing modest power improvement in detecting light-tailed misspecifications, PITOS otherwise performs comparably to POT-C but suffers from Type I error miscalibration: overly conservative for continuous model cases, and overly liberal for discrete data models (top row).}
 \label{fig:pitos}
\end{figure}

The PITOS method developed in \citet{covington2025powerful} extends POT-C by supplementing marginal order statistic tests with conditional tests between order statistic pairs. In addition to computing $p$-values from the marginal beta distributions of individual order statistics \(u_{(j)}\), PITOS also computes p-values from the conditional (beta) distributions of \(u_{(j)}|u_{(i)}\), Eq.~\eqref{cbeta}, for pairs \((i,j)\in \{1,\ldots,n\}\). The p-values are then aggregated using the Cauchy combination technique.  Index pairs $(i,j)$ are selected to evenly cover the unit square \([0,1]^2\) via their scaled coordinates \((i/n, j/n)\). In theory, PITOS's inclusion of pairwise conditional structure should enable detection of localized departures from the null that marginal-only tests may miss.
\begin{equation}
    F_{u_{_{(j)}}|u_{_{(i)}}}=
\begin{cases}
\displaystyle  \Beta\left((u_{_{(j)}}-u_{_{(i)}})/(1-u_{_{(i)}})\mid j-i,n+1-j\right), & i<j, \\
\Beta\left(u_{_{(j)}}/u_{_{(i)}}\mid j,i-j\right), & i>j.
\end{cases}
\label{cbeta}
\end{equation}
However, when testing dependent PIT values (e.g., LOO-PIT), aggregating PITOS's much larger collection of p-values increases the risk of poor Type I error control. The pair generation may also produce duplicate $(i,j)$, introducing perfectly correlated tests that amplify conservativeness
beyond what the 1.15× correction addresses, even with increasing data size ( see Figure~\ref{fig:pitos}). On the other hand, conditional distributions \(  F_{u_{_{(j)}}|u_{_{(i)}}=x}\) require the actual value \(x\), not just its rank, making PITOS sensitive to ties, and numerical precision. With weighted rank LOO-PIT, values may include exact $0$'s and $1$'s and inter-observation spacing becomes imprecise. This discretization can lead to problematic conditioning events where the conditional beta CDF evaluations fails numerically.

To summarize, PITOS will be more influenced by dependence and discretization artifacts than the marginal-only POT-C approach. We recommend restricting the PITOS procedure to continuous data models with exact probability integral transforms. In the examples explored in this paper, we have not seen any significant benefit for PITOS from the increased computation time needed for this more elaborate test. See Figure~\ref{fig:pitos}.

\end{document}